\documentclass[10pt]{article} %can change fontsize

\usepackage{amsmath,amsfonts,amssymb,amsthm}
\usepackage{mathtools}
\usepackage{graphicx,float}

\usepackage{fullpage}
\usepackage[top=1.9 cm, bottom = 1.9 cm, left = 1.9 cm, right = 1.9cm]{geometry}
\usepackage{cancel}
\usepackage{mmacells}
\usepackage{systeme}

\DeclareMathOperator\sech{sech}

\makeatletter

\newcommand{\Rmnum}[1]{\expandafter\@slowromancap\romannumeral #1@}
\makeatother

\newcommand*{\bpm}{\begin{pmatrix}}
\newcommand*{\epm}{\end{pmatrix}}

\newcommand{\bg}[1]{\begin{gather*}#1\end{gather*}}

\newcommand{\bmc}[2][y]{\begin{mmaCell}[index=#2,functionlocal=#1]{Code}}
\newcommand*{\bmco}{\begin{mmaCell}{Output}}

\mmaDefineMathReplacement{Lambda1}{\lambda_1}
\mmaDefineMathReplacement{Lambda2}{\lambda_2}
\mmaDefineMathReplacement{-Lambda1}{-\lambda_1}
\mmaDefineMathReplacement{-Lambda2}{-\lambda_2}
\mmaDefineMathReplacement{weirdx}{\times}
\mmaSet{moredefined={ArrayFlatten, J, H, L,Removeboxes}}

\mmaSet{
  morefv={gobble=0},
  linklocaluri=mma/symbol/definition:#1,
  morecellgraphics={yoffset=1.9ex}
}
\usepackage{bbold} %for indicator function 
\newcommand*{\citealt}[1]{\cite{#1}}
\newcommand*{\citep}[1]{(\cite{#1})}
\theoremstyle{plain}
\newtheorem{theorem}{Theorem}
\newtheorem{corollary}{Corollary}
\newtheorem{prf}{Proof}

\theoremstyle{definition}
\newtheorem{definition}{Definition}

\title{Fast Calibration of Car Following models to Trajectory data using the Adjoint Method}
\author{Ronan Keane \\  Systems Engineering, Cornell University, Email: (rlk268@cornell.edu) \\ 
H. Oliver Gao \\ Civil and Environmental Engineering, Cornell University,  Email: (hg55@cornell.edu)}

\begin{document}
\maketitle

\section*{Abstract}
Before a car-following model can be applied in practice, it must first be validated against real data in a process known as calibration. This paper discusses the formulation of calibration as an optimization problem, and compares different algorithms for its solution. The optimization consists of an arbitrary car following model, posed as either an ordinary or delay differential equation, being calibrated to an arbitrary source of trajectory data which may include lane changes. \\
\indent Typically, the calibration problem is solved using gradient free optimization. In this work, the gradient of the optimization problem is derived analytically using the adjoint method. The computational cost of the adjoint method does not scale with the number of model parameters, which makes it more efficient than evaluating the gradient numerically using finite differences. Numerical results are presented which show that quasi-newton algorithms using the adjoint method are significantly faster than a genetic algorithm, and also achieve slightly better accuracy of the calibrated model. 

\pagebreak

\section{Introduction}\label{intro}
\subsection{The Case for Car Following and Trajectory Data}\label{intro1}

A complete traffic simulator consists of many different modules, including algorithms to determine the demand for different road sections, complex rules to govern lane changes, mergers, and diverges, a model to describe traffic flow and driving behaviors, and more. Out of all these modules, the description of traffic flow is the most fundamental. The decision of what model is used to describe traffic flow has impacts on the simulator's strengths, weaknesses, and overall performance. In particular, one important distinction is whether to simulate at the macroscopic or microscopic level. 

At the microscopic level, car following models are often used. Car following models describe how a vehicle behaves as a function of its surroundings; typically they are expressed as differential equations where any vehicle is coupled to the vehicle directly ahead (i.e. its lead vehicle). Many car following models are based on common sense rules of driving. Some simple examples are adopting a speed based on the distance between you and your leader, or accelerating based on the difference in speeds between you and your leader. 

Whereas microscopic traffic models describe the motions of individual vehicles, their macroscopic counterparts treat traffic as a fluid. At the macroscopic level, there are no vehicles, only traffic densities. Congested road sections with high densities slowly trickle forward, while uncongested traffic at low densities flow freely. This connection to fluid dynamics was originally made by \cite{39}, and the influence of the resulting traffic flow theory can hardly be overstated. In the more than six decades since, numerous works have attempted to improve upon the so-called Lighthill-Whitham-Richards (LWR) model in order to better describe various complex elements of traffic flow. Among those, \cite{37}, \cite{40}, \cite{46}, \cite{48}, \cite{53} are perhaps some of the most notable. Nonetheless, when it comes to macrosimulation, the LWR model, and its discrete numerical formulation \citep{42}, is still widely used today. 

For all its merits, the LWR model is not without its limitations. A key relationship between flow and density known as the ``fundamental diagram", must be supplied. More importantly, it is assumed vehicles are homogeneous and exhibit static behavior.
In real traffic, drivers behave differently; vehicles have varying performances; traffic demand fluctuates; controlled intersections and lane changes have complex effects. Traffic flow is incredibly dynamic and inhomogeneous. 
The shortcomings of the LWR model become apparent in practice (see \cite{46,49,41}). In the fundamental diagram, instead of equilibrium curves, one finds a wide scattering of points. The time evolution of real traffic flow shows a much richer picture than what the LWR model predicts. Examining the evolution of a vehicle trajectory in the speed-spacing plane, one sees complex hysteresis loops which are inconsistent with current macroscopic theory (\citealt{28}, \citealt{50} or Fig. \ref{fig2} are examples of such plots). Given these limitations, one would hope for some car following model that could explain all these phenomena.   

However, existing works are not prepared to offer guidelines as to which car following models are best. It is also not clear as to how much more accurate car following models are compared to macro-models (like LWR). Answering either of these pressing questions is difficult: the traffic microsimulation literature consists of literally hundreds of different models, many of which have never actually been thoroughly validated \citep{7}. Works such as \cite{9}, \cite{10}, \cite{14}, and \cite{19} have attempted to address this long-standing problem of car following models but have had varying results as to which models are best. 

Drawing conclusions on the efficacy of different car following models requires both a large vehicle trajectory dataset (e.g. \citealt{6} or \citealt{66}) to facilitate comparison, as well as a reliable methodology for model calibration and validation. The aim of this work is to develop new methodology to improve the calibration of car following models. However, one should keep in mind that this only addresses one half of the challenge; equally important is the creation of new datasets and new approaches for collecting accurate and meaningful traffic data. 

There is a need for an increased use of trajectory data to study traffic flow. This need goes far beyond the validation of existing car following models. As new technologies like autonomous vehicles, connected cities and on-demand mobility become fully realized, the LWR model assumption of homogeneous drivers exhibiting static behavior becomes increasingly unrealistic. The limitations of macrosimulation and the uncertainty of new technology are both good reasons to suspect a growing demand for microsimulation. At the same time, vehicle trajectory data (which is needed for microsimulation/car following) is easier to get than ever before because of new technologies like computer vision, lidar/radar, and GPS. Compared to the more commonly used loop detector data, trajectory data describes traffic flow at much higher temporal and spatial resolutions. These new sources of data, still largely unexploited, can help advance both micro- and macro- theory. 

\begin{figure}[h] 
\centering 
\includegraphics[ width=\textwidth]{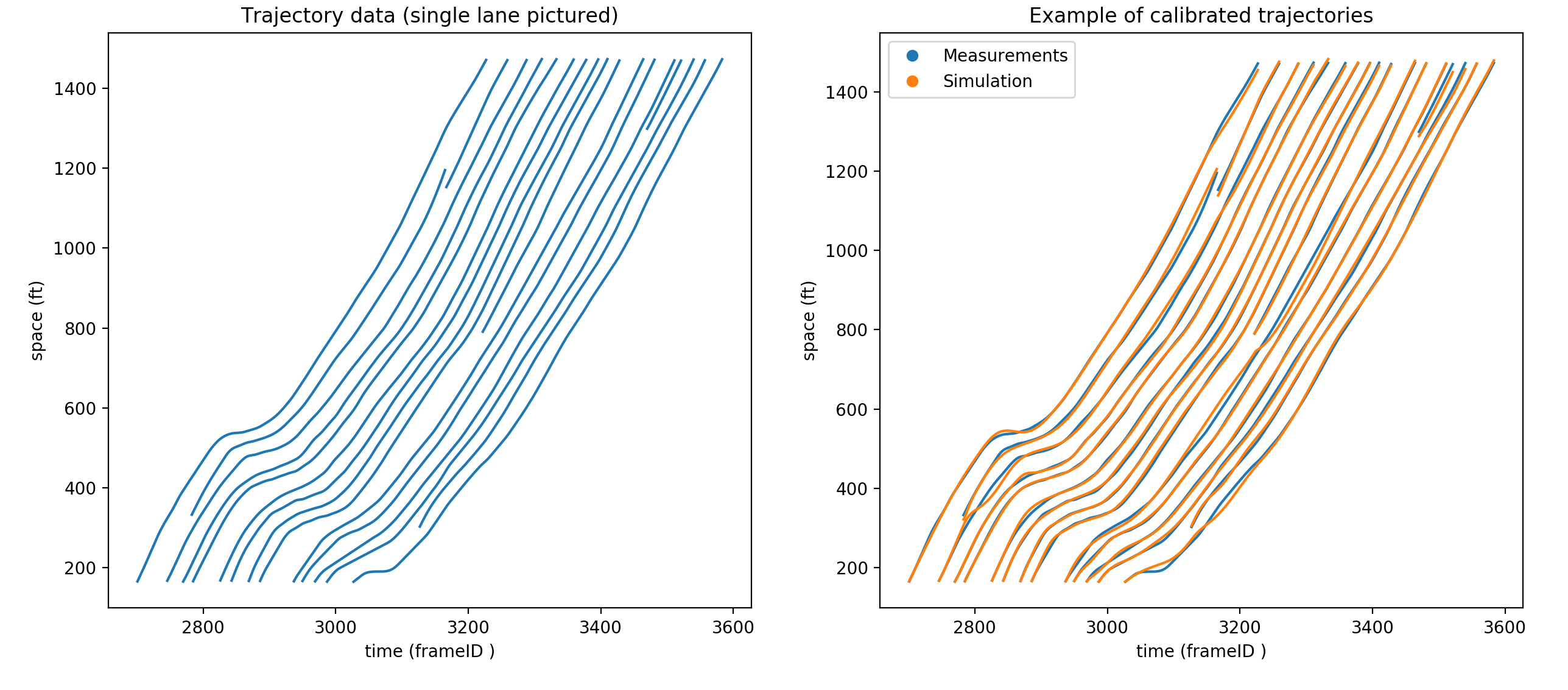} 
\caption{On the left, an excerpt of a single lane from the NGSim data. ``Incomplete" trajectories are due to vehicles changing to/out of the lane pictured. One can observe many of the important features of traffic flow such as shockwaves, lane changing events, and individual driver behaviors. Compared to macroscopic data, these features are captured at higher temporal/spatial resolutions and without aggregation. On the right, a car following model (the OVM) has been calibrated to the trajectory data.   } \label{fig1}
\end{figure}

\subsection{Calibration Using Trajectory Data}\label{intro2} 
\noindent It is relatively simple to write down some mathematical equations and call them a traffic model. Far more difficult is applying this model in practice. In order to validate a model against real data, one must find the model parameters that best explain the data: this is known as calibration. Solving the calibration problem requires both a dataset describing the quantity of interest, as well as a mathematical model that explains the data. This paper deals with car-following models that are posed as differential equations, and may include elements such as delays, bounds on acceleration/speed, and multiple driving regimes. The car-following model is to be calibrated against trajectory data, time series data that describes the motion of individual vehicles. 

Existing literature has covered various aspects of the trajectory data calibration problem: the use of meta-models, sensitivity analysis, optimization algorithms, measures of performance (e.g. speed or spacing data), and goodness-of-fit function (i.e. loss function). \cite{treibercalibration}, \cite{24}, and \cite{25} are good works focused primarily on methodology. \cite{1}, \cite{12}, \cite{14} and \cite{19} are good works focused on both calibration and the benchmarking of various models. 

In this paper calibration is posed as an optimization problem. For a wide class of models, being calibrated to trajectory data gathered from any source, the gradient of the optimization problem has been derived in an efficient way using the adjoint method. This enables the calibration problem to be solved by more efficient optimization routines which use gradient information. Several state of the art algorithms are benchmarked in order to determine both the accuracy of the fit they can achieve, as well as the computational effort required to apply the algorithm. The paper is organized as follows. Section \ref{models} introduces car following models and discusses current research directions for improving car following models. Section \ref{calibration} describes the optimization problem and discusses some issues arising in its solution. Section \ref{adjoint} applies the adjoint method to the optimization problem and compares it to finite differences. Section \ref{algos} compares different algorithms for solving the calibration problem.

\section{Car-Following Models}\label{models}

\noindent We will consider any car-following model that can be expressed as either an ordinary or delay differential equation. This describes the majority of models in the literature, as well as the majority of models used in commercial microsimulation packages \citep{47}. We deal with models of the following form 
\[ \ddot  x_n(t) = h(\dot x_{n-1}(t-\tau), x_{n-1}(t-\tau), \dot x_{n}(t-\tau), x_{n}(t-\tau), p)  \tag{1}\label{1}\]
where $x_n(t)$ is the position of vehicle $n$ at time $t$. $h$ is the car-following model, which depends on parameters $p$, the ``following" vehicle $x_n$ and ``lead" vehicle $x_{n-1}$. $\tau$ is a time-delay parameter which represents human reaction time.  Eq. \eqref{1} states that a driver's behavior (how they choose to accelerate) is determined from the speed and position of their own car, the speed and position of the leading car, and parameters $p$. Given a lead vehicle trajectory $x_{n-1}(t)$, parameters $p$, and initial condition $x_n(t_n)$, the car following model $\eqref{1}$ will produce the follower trajectory $x_n(t)$. 
The presence of nonzero $\tau$ leads to a delay differential equation (DDE).  Many car-following models (including most of those used in commercial software) treat reaction time explicitly by using a DDE. 

Besides ODEs and DDEs, some car-following models are formulated as stochastic differential equations or derived using nonparametric techniques. We briefly discuss these types of models in section \ref{sde} but the current methodology does not address their calibration.

\subsection{Ordinary Differential Equations}\label{ODE}

\noindent Eq. $\eqref{1}$ gives a general form for car-following models. The simplest type of model is a first-order ODE where the speed of the following vehicle $n$ is completely determined by the space headway $(x_{n-1} - x_n)$. In this type of model, each driver adopts a speed based on the distance between their car and the leading car. 

This simple kind of model is important because of its connection to macroscopic theory and the LWR model. Specifically, if one assumes all vehicles are identical, and interprets the reciprocal of density as headway, then a functional relationship between traffic flow and density can be recast as a relationship between vehicle speed and headway \citep{20}. Under these assumptions, the fundamental diagram can be thought of as describing a first-order car-following model. Consider the Newell car following model an example
\[\dot x_n(t) = \min\{ v_f, \dfrac{x_{n-1}(t) - x_n(t) - s_{\text{jam}}}{\alpha}\}. \tag{2}\label{2} \]
$v_f, \alpha$  and $s_{\text{jam}}$ are parameters describing a triangular fundamental diagram \citep{1}. Namely $v_f$ is the speed of a vehicle during free flow, $s_{\text{jam}}$ is the distance between two completely stopped vehicles, and $-s_{\text{jam}}/\alpha$ is the shockwave speed. 

The Newell model can also be written as a trajectory translation model, where the trajectory of a following vehicle is equivalent to the lead vehicle trajectory, translated over both space and time.
\[ x_n(t) = x_{n-1}(t - \alpha) - s_{\text{jam}}  \tag{3}\label{3}\]

\begin{figure}[h] 
\centering 
\includegraphics[ width=\textwidth]{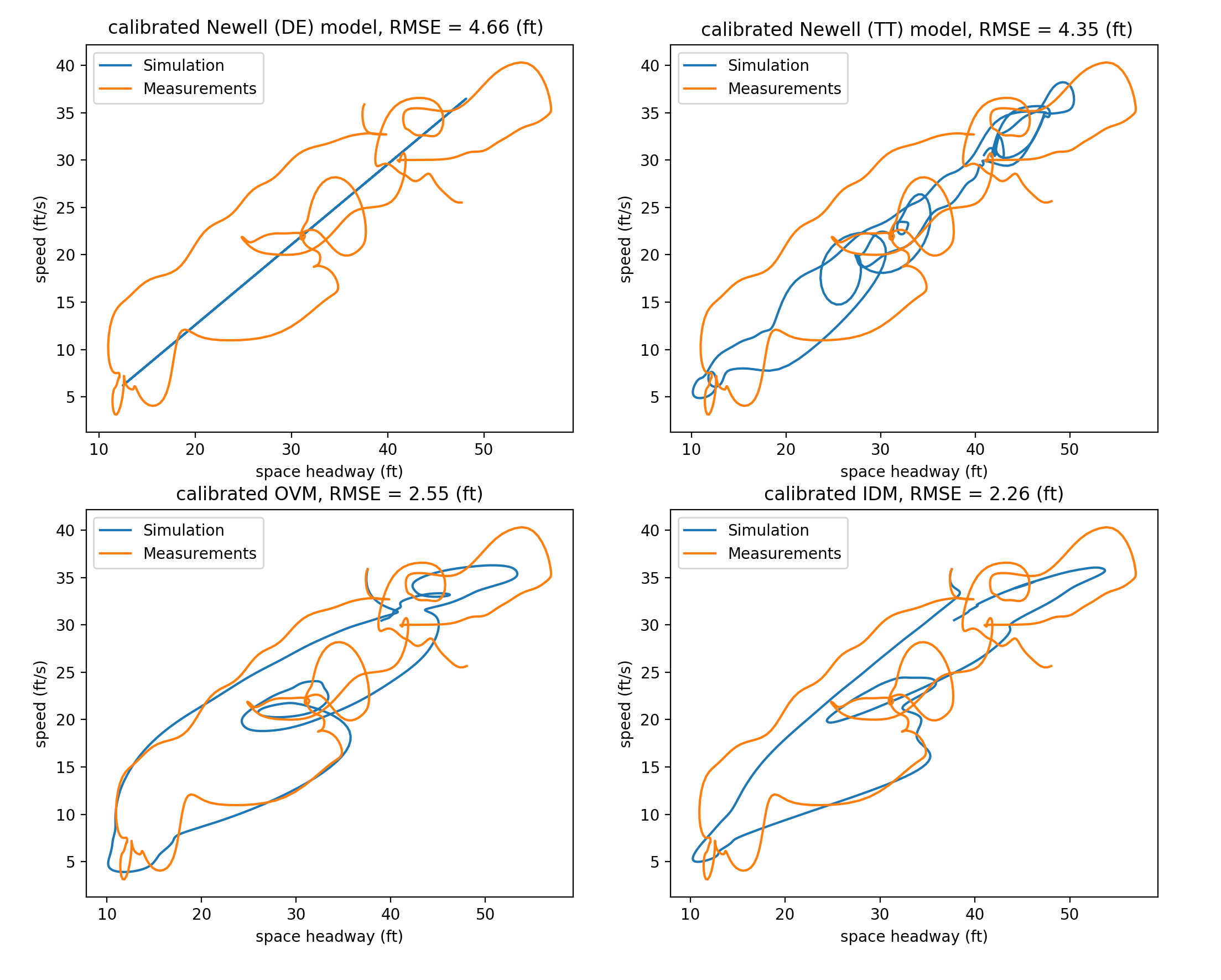} 
\caption{Four different car following models are calibrated to a single empirical vehicle trajectory by minimizing the root mean square error between simulation and measurements. The trajectories are plotted in the speed - headway plane.} \label{fig2}
\end{figure}

\noindent We refer to Eq. \eqref{3} as Newell (TT) and Eq. \eqref{2} as Newell (DE), where the acronyms stand for trajectory translation and differential equation, respectively. In the special case where the TT and DE models both use a numerical time step equal to $\alpha$, and vehicles are always in the congested regime (i.e. not moving at their free flow speeds), the two models become equivalent \citep{20}. In this special case, either model gives trajectories that exactly obey the LWR model with a triangular fundamental diagram. In general however, the two models are different (see for example Fig. \ref{fig2}). The advantages of the Newell (TT) or (DE) models are that they contain a small number of parameters, and those parameters are easily interpretable under macroscopic theory. 

More complicated car-following models (compared to those of Newell) are sometimes criticized for two main reasons: first, for not having enough empirical support for their validity, and second, for having too many parameters. Despite these criticisms, more complicated models have behaviors that attract researchers. For example, the optimal velocity model (OVM) proposed by \cite{52} exhibits a spontaneous phase transition between free flow and congested states, which may describe ``phantom" traffic jams \citep{54}. Another example is the intelligent driver model (IDM) proposed by \cite{69}, which can qualitatively reproduce a number of complex (empirical) traffic states. 
Besides more recent models like OVM or IDM, there are also a plethora of older models to complicate the mix \citep{7}. The default models in most modern traffic simulator packages (for example, VISSIM or AIMSUN) are still older car-following models \citep{47}, presumably because it is unclear if the newer car-following models offer any benefits. 

Fig. \ref{fig2} compares each of the four models just mentioned by calibrating them to a single vehicle trajectory from the reconstructed NGSim data \citep{29}. The calibration minimized the root mean square error (RMSE) between the simulation and measurements. The vehicle (ID 1013) was recorded for only 55 seconds, but even in that time many hysteresis loops were recorded in the speed-headway plane. For the NGSim data, which was recorded on the freeway in congested conditions, trajectories that qualitatively resemble this one are the rule rather than the exception. The different models all give a different caricature of the oscillatory behavior observed. In particular, the Newell (TT) model (with a simulation time step of .1 s) appears to better capture the seemingly chaotic nature of the oscillations, whereas the more complicated (and computationally expensive to calibrate) IDM or OVM models result in a lower RMSE and better capture the width of the hysteresis (the differences in speeds at a given headway). The point of this figure is not to make a case for one model over the other, but rather to point out the interesting behavior shown even in a single short trajectory. It also raises an important question which existing research seems unprepared to answer. 

With so many different models, each of varying complexities, which ones best describe traffic flow? Our work on calibration is motivated by this question.

\subsection{Delay-Differential Equations} 
\noindent Compared to ODEs, DDEs are more complicated to analyze and solve. Car-following models typically only consider the simplest case, where time delay (reaction time) is a constant. It seems the distinction between ODEs and DDEs has been given limited importance in the calibration literature, even though the addition of time delay as a parameter causes extra complications. As for the literature as a whole, the effect of nonzero time delay on car-following models is still not fully understood.  

When studying the behavior of car-following models, one common approach is to consider a line of cars with identical model parameters on a closed loop. In this special setting, one can find an ``equilibrium" solution where all vehicles have the same speed and same headway, with zero acceleration. The equilibrium solution is contrasted with an oscillatory/periodic solution, where the speed of a vehicle can oscillate in a manner reminiscent of stop-and-go or congested traffic. Since most analytic (i.e. not numerical) results come from studying the behavior around the equilibrium solution, the stability of the equilibrium solution is particularly important. Adjusting parameters in a model can be said to have a stabilizing (destabilizing) effect if the equilibrium solution will persist (degenerate) under a wider variety of conditions. Intuitively, one expects that drivers having slower reaction times will have a destabilizing effect, and this seems to be generally correct. 

 It has been shown for both the optimal velocity model (OVM) and intelligent driver model (IDM) that the inclusion of time delays has a destabilizing effect  (see \citealt{16}, \citealt{56}, \citealt{5}). But time delay is not the only model parameter that can have such effects. For example, \cite{57} showed that increasing (decreasing) the OVM's ``sensitivity parameter"  has stabilizing (destabilizing) effects (the sensitivity parameter is proportional to how strongly the driver accelerates). Time delay can also have more subtle effects: \cite{56} found the inclusion of time delays for OVM caused bistability between equilibrium and oscillatory solutions.   

 In \cite{15} the OVM is considered with different reaction times $\leq 0.3 $ seconds. In this range, no qualitative changes to the model are observed. The authors state that changing the sensitivity parameter can account for the effects of reaction time. \cite{12} and \cite{16} have done experiments on OVM with larger reaction times ($\approx 1$ seconds) and found that while it is true that reaction time is not especially important at low values, once increased past some threshold it will start to cause unstable behavior leading to vehicle collisions. A similar behavior was found in \cite{23} for the Gipps car-following model. These findings show the inclusion of reaction time will have a destabilizing effect, but large reaction times seem to have especially unstable effects, perhaps owing to some sort of phase transition. 

 In practice, one can observe that calibrated values for reaction time greatly vary across the literature, even for the same model. Consider that \cite{58},\cite{14}, and \cite{19} all calibrate the Gipps model, but all find very different means and variances for drivers' reaction times. One found the mean to be 0.57 s with a variance of 0.02 s. Another found the mean to be 1.73 s with variance 1.35 s.  
 
 Besides reaction time, another interesting extension to a model is the inclusion of ``anticipation" effects. Examples of anticipation are a driver estimating the future speed/position of a leading vehicle, or a following vehicle reacting to multiple vehicles instead of just its leader. \cite{5} showed that anticipation has a stabilizing effect for IDM, and \cite{21} showed the same for OVM. Of course, due to nonlinearities, stabilizing and destabilizing effects will not simply cancel out. 
\subsection{Stochastic and Data-Driven Models} \label{sde}
In this paper we consider the calibration of car following models formulated as ODEs or DDEs. This section discusses some recent research directions in which car following models are formulated as stochastic differential equations (SDEs), or derived from data using nonparametric models. \\
Papers using SDEs to model traffic flow typically argue that the stochastic models are better able to describe the formation and growth of congestion. Besides being able to achieve a closer fit to data, authors have concluded that deterministic models cannot reproduce some qualitative features of congestion. In particular, some recent field experiments involving 25 and 51 human driven vehicles on a straight road \citep{experimentalandempirical} showed that the standard deviation of traffic speeds initially grow in a concave way, whereas deterministic models seem to only be able to show convex growth. \\
SDEs in the literature are usually based on modifying an existing deterministic ODE model rather than created as an entirely novel model. Two common stochastic mechanisms are acceleration noise, and action points. For acceleration noise, white noise is added to the output of the deterministic model. The white noise may be sampled from either a uniform or normal distribution. Closely related to acceleration noise, is the case where the noise is added to a model parameter instead of directly to the output of the model. In \cite{laval2014} and \cite{flowarchitecture} gaussian acceleration noise was added to the Newell model and IDM respectively. In 2D-IDM from \cite{empiricalanalysis} uniform noise is added to a model parameter (in that case, the desired time headway parameter). \\
Action points are implemented by creating a model regime where the driver is insensitive to changes in the headway and/or lead vehicle velocity. Thus when inside the regime, a driver will maintain their current behavior until a sufficiently large difference in the state is achieved, at which point the driver will react to the now significantly different driving situation. The reader is referred to \cite{treiber2017} or \cite{onsomeexperimental} for specific details on implementing action points. \\ 
The motivation for including white noise is to model random human behavior and reflect the fact that humans cannot perfectly control their vehicles, whereas the motivation for action points is that drivers are not able to perceive minute changes to the traffic situation and thus a sufficiently large difference is required before they can start to react. \cite{treiber2017} shows that action points and white noise have similar effects on traffic simulations, and suggests that both mechanisms should be included. \\

\noindent Data-driven car following models are based on machine learning models and algorithms, and as a result their formulation is significantly different than the ODE/DDE/SDE discussed so far. However, despite their different formulation, the inputs (current headway, follower and leader velocities) and output (current follower acceleration) are typically very similar. Papers in the literature have reported that data-driven models can achieve better fits to empirical data compared to parametric models such as OVM or IDM. The discussion of these machine learning techniques is outside the scope of this paper but we mention the different approaches which have been successfully applied. In \cite{simplenonparametric} $k$-nearest neighbor (knn) is used, and \cite{towardsdatadriven} uses locally weighted regression (loess), which can be thought of as a generalization of knn. \cite{recurrentcf} discuss using feedforward neural networks with a single hidden layer and ReLU nonlinearity. That paper also uses recurrent neural networks (RNNs) to overcome some difficulties of the feedforward neural networks. \cite{lstmcf} uses long short-term memory neural networks, a special kind of RNN. \cite{deeprlcf} use deep reinforcement learning (deep RL) and policy gradients, and do a comparison between deep RL, IDM, RNN, loess, and feedforward neural networks.

\section{Calibration}\label{calibration}
\noindent It is useful to rewrite the second order differential equation Eq. $\eqref{1}$ as a system of first order equations:
\[ \dot x_n (t) = \bpm \dot x_n^*(t) \\ \dot v_n(t) \epm = \bpm v_n(t) \\  h^*( x_{n-1}(t-\tau_n), x_{n}(t-\tau_n), p_n)  \epm = h_n(x_n(t), x_n(t-\tau_n), x_{L(n)}(t-\tau_n), p_n)  \tag{4}\label{4}\]
\noindent where $x_n$ is now a vector consisting of the position and velocity of vehicle $n$ at time $t$. $x_n^*$ and $v_n$ refer to the position and speed for vehicle $n$. $p_n$ denotes the parameters for vehicle $n$. $L(n,t)$ refers to the leader of vehicle $n$ at time $t$, and we denote the lead trajectory as $x_{L(n)}$ (time dependence suppressed) to reflect the fact that the index corresponding to the lead vehicle may change due to lane changing. Eq. \eqref{4} is what we refer to when we say ``car-following model" and its solution $x_i(t)$ is a ``simulated trajectory".

\noindent The calibration is done for some arbitrarily sized platoon of $n$ vehicles, indexed as $1, \ldots, n$. Each vehicle has some individual parameters $p_i$ and a corresponding car following model $h_i$, so all vehicles have their own set of parameters, and $p_i$ only describes vehicle $i$. Let $p$ be a concatenated vector of all the parameters, so $p = [p_1, p_2, \ldots, p_n]$. Calibration finds the parameters $p$ which minimize the total loss between the simulated trajectories $x_i(t)$ and the corresponding measured vehicle trajectories, which are denoted $\hat x_i(t)$. The loss function is denoted as $f(x_i, \hat x_i, t)$ and represents the goodness of fit of the simulated trajectory $x_i$ at time $t$. The measurements $\hat x_i(t)$ are known in the interval $[t_i, T_i]$ for each $i$. \\
In order to compute the simulated trajectories for each vehicle $i$, we require a lead vehicle trajectory $x_{L(i)}(t)$ as well as an initial condition $x_i(t_i)$. The time interval that the lead vehicle trajectory is known is denoted $[t_i, T_{i-1}]$. 
$x_i(t_i)$ is the initial condition, which is specified as $x_i(t_i) = \hat x_i(t_i)$ for an ODE model. 

For an ODE, the optimization problem is 
\begin{align*}
\underset{p}{\min} & \quad  F = \sum_{i=1}^n\int_{t_i}^{T_{i-1}} f(x_i, \hat x_i, t) dt          \tag{5}\label{5} \\
\text{s.t.} & \quad   \dot x_i(t)  - h_i(x_i(t),x_{L(i)}(t),p_i)  = 0, \quad t \in [t_i, T_{i-1}], \ \ \ i = 1, \ldots, n   \\
& \quad     x_i(t_{i}) - \hat x_i(t_{i})= 0 \quad  i = 1, \ldots, n    \\
& \quad b_{\rm low} \leq p \leq b_{\rm high}
\end{align*}
The objective function $F$ is defined as the integral of the loss function $f$ over time, summed over each simulated vehicle. The constraints are that the simulated trajectories $x_i$ obey the car-following model $h_i$ from time $t_i$ to $T_{i-1}$, with the initial conditions $x_i(t_{i}) = \hat x_i(t_{i})$. There may also be some box constraints $b_{\rm low}, b_{\rm high}$ which prevent parameters from reaching unrealistic values.  

The DDE case is similarly defined: 
\begin{align*}
\underset{p}{\min} & \quad  F = \sum_{i=1}^n\int_{t_i+\tau_i}^{T_{i-1}^*} f(x_i, \hat x_i, t) dt          \tag{6}\label{6} \\
\text{s.t.} & \quad   \dot x_i(t)  - h_i(x_i(t), x_i(t-\tau_i),x_{L(i)}(t-\tau_i),p_i)  = 0, \quad t \in [t_i+\tau_i, T_{i-1}^*], \ \ \ i = 1, \ldots, n   \\
& \quad     x_i(t) -\hat x_i(t) = 0 \quad t \in [t_ i, t_i+\tau_i], \ \ \ i = 1, \ldots, n    \\
& \quad b_{\rm low} \leq p \leq b_{\rm high}
\end{align*}
where $T_{i-1}^* = \min\{T_{i-1}+\tau_{i}, T_i\}$, and the reaction time for vehicle $i$, $\tau_i$, is assumed to be a model parameter. 
$x_i(t) -\hat x_i(t) = 0$ defines the history function for $x_i$ (DDEs have history functions as their initial conditions). 

\subsection{Downstream Boundary Conditions for Car Following Models} \label{downstreamboundary}
\begin{figure}[H] 
\centering 
\includegraphics[ width=.6\textwidth]{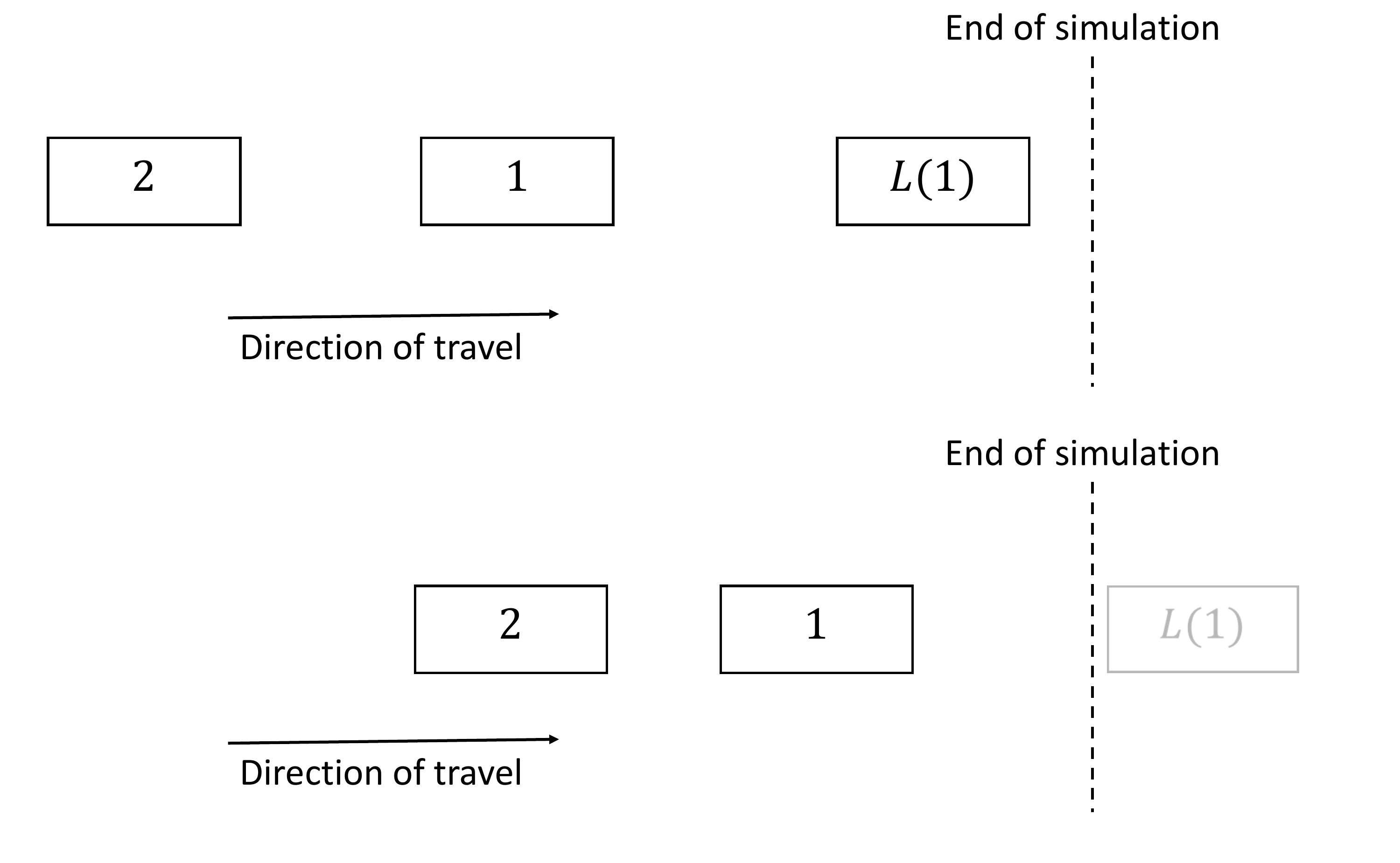} 
\caption{Boxes represent vehicles; vehicle $2$ follows $1$, and $1$ follows $L(1)$. In top panel, the leader $L(1)$ is still in the simulation and so the car following model can be used to update vehicle $1$. In bottom panel, $L(1)$ is no longer in the simulation, and so a downstream boundary condition is used to update vehicle $1$. }  \label{boundary}
\end{figure}

In Eqs. \eqref{5} and \eqref{6} we have specified the initial conditions, which give the times vehicles enter the simulation and their position/speed upon entering. These conditions might also be referred to as upstream boundary conditions since they state how vehicles enter the simulation. For video data such as NGSim, which is recorded on a fixed section of road, downstream boundary conditions also need to be specified. Video data has staggered observation times ($T_{i-1} < T_i$) meaning that a vehicle's leader leaves the study area before the vehicle itself. Figure \ref{boundary} gives a visual depiction of this. \\
Car following models require a lead vehicle trajectory as input, therefore once the lead vehicle leaves the simulation, the car following model can no longer be used. Downstream boundary conditions then give a rule for the dynamics a vehicle follows once its lead vehicle leaves the simulation (i.e. they specify the dynamics for $x_i(t), \ \ t \in [T_{i-1}, T_i]$). The the boundary conditions we implement in this paper are
\begin{align*} 
\dot x_i(t) = \dot{\hat{x}}_i(t), \ \ t \in [T_{i-1}, T_i] \tag{7}\label{7}
\end{align*}
meaning that when a vehicle's leader is no longer available, the vehicle will move with whatever speed was recorded in the data. Under these boundary conditions, the velocity $v_i(t)$ and acceleration $\dot v_i(t)$ will be piecewise continuous (they will have a jump discontinuity at $T_{i-1}$). The objective $F$ and gradient $dF/dp$ are still continuous under this boundary condition (see sections \ref{calibration8} and \ref{continuous} for the relevant discussion on discontinuities in car following models and how they impact the calibration problem). Note that for a DDE, the time $T_{i-1}^*$ is used instead of $T_{i-1}$.  \\
It is necessary to specify the downstream boundary conditions for two reasons. First, it allows vehicles to exit the simulation area. Second, it allows congestion to propogate into the simulation area from downstream. This happens when the boundary conditions specify a reduction in speed.
\subsection{Reaction Time as a Parameter}\label{calibration6}
Simulated trajectories are computed numerically with a specified time discretization. Since we want to compare the simulated trajectories with the actual measurements, it makes sense for the simulation to use the same time discretization as the data. For example, if the data is recorded every 0.1 seconds, the simulated trajectories should use a time step of 0.1 seconds.  If the simulation used a different time step (e.g. 0.09), then the loss can only be calculated at a few points (e.g. 0.9, 1.8) unless we define some interpolation procedure. 

For a DDE, the time discretization and reaction time need to be multiples of each other. If the simulation uses a time step of 0.1 seconds, then the reaction time will have to be an integer multiple of 0.1 (e.g. 0.1, 0.2). We can either: 
\begin{enumerate} 
\item Enforce that every vehicle's reaction time is a integer multiple of 0.1 
\item Allow reaction time to be a continuous variable and define the loss function through interpolation. %
\end{enumerate}
If we choose 1, it leads to a mixed integer nonlinear programming problem (MINLP). This is a much harder problem than the original nonlinear program. One approach used in \cite{19} would be to split up all vehicles into small platoons (either pairs or triplets) and use a brute force search for reaction time. However, even when the platoons are small, using brute force vastly increases computation because the calibration has to be repeated for every possible combination of reaction times. 

The other option is to use an approach similar to \cite{12} where reaction time is a continuous variable and interpolation is used on the simulated trajectories. In general, any sort of interpolation could be used, but linear interpolation is the most common. In this approach, interpolation needs to be used on the lead trajectory and initial history function in order to define the model. Assuming the numerical time step is not equal to the delay, interpolation will need to be used on the simulated trajectory as well. 

One issue with interpolation is how to compute the loss function. Specifically, assuming the simulation is defined on a different set of times than the measurements, should the simulation be interpolated onto the measurement's times or vice versa? We suggest doing both. This avoids the possibility of error cancelation occurring during the interpolation and causing the loss function to be erroneously small. In this scenario it would be preferable to reweight the loss function as well since the loss will be computed over more points. 

\subsection{Multiple Regimes, Lane Changing, and Discontinuities in Car Following Models}\label{calibration8}
We say a model has multiple regimes when the model is expressed either as a piecewise function, or by using $\min$ and $\max$ functions. These types of modifications are frequently applied in practice. Perhaps the most common are examples like Eq. \eqref{2}, where the maximum speed is explicitly constrained by some constant $v_f$. For second order models, often the acceleration will be constrained between minimum and maximum accelerations $a_{\rm min}, a_{\rm max}$. Besides these simpler regimes which implement bounds on the acceleration and/or velocity, there are also models which use different regimes to describe different car following behavior. The Gipps model \citep{59} is one such example, which has one regime to describe normal car following behavior, and a second regime which implements an emergency braking behavior. Other examples of multi-regime models are the constant acceleration heuristic proposed in \cite{treiberbook}, the follower stopper controller proposed in \cite{FS}, or any model implementing action points. We consider 
\begin{align*} 
h_i(x_i, x_{L(i)}, p_i) = \begin{cases} 
 h_{i,2} & g_{i,2}(x_i, x_{L(i)}, p_i) \geq 0  \\ 
  h_{i,1} & \text{ otherwise} \\ 
  \end{cases} \tag{8}\label{regime}
\end{align*}
as a general form of a two regime car following model, with the two regimes $h_{i,1}$ and $h_{i,2}$. The condition for the associated regime to become activated is $g_{i,2}$ (which is possibly a vector valued function). For example with Eq. \eqref{2}: $h_{i,1} = (x_{L(n)} - x_n - s_{\rm jam})/\alpha$, $h_{i,2} = v_f$ and $g_{i,2} = h_{i,1}   - h_{i,2}$. There must always be $m-1$ such conditions for a model with $m$ regimes, so a model with more regimes will simply have $h_{i,2}, \ldots, h_{i,m}$ and $g_{i,2}, \ldots, g_{i,m}$. \\
In the case of a first order model's velocity being bounded, or a second order model's acceleration being bounded, the resulting model $h_i$ is everywhere continuous (because both regimes will be equal when the switching condition $g_{i,2}$ is met). In general a multi-regime model need not be everywhere continuous. Gipps and follower stopper are both examples of a multi-regime model being everywhere continuous; the constant acceleration heuristic or a model implementing action points are examples not guarenteed to be continuous when switching between regimes. Of course, even for a multi-regime model which is everywhere continuous, its derivatives will have discontinuities when switching between regimes. Note that a multi-regime model written using $\min$ and $\max$ will always be continuous. The reason why we prefer using the form Eq. \eqref{regime} is because 1) it is more general, 2) we can explicitly refer to the switching condition $g_{i,2}$ and 3) the regime of the model is explicitly stated, which avoids the problem of the regime being ambiguous when the quantities inside a $\min / \max$ are equal. \\
Lane changing is another issue which causes discontinuities in car following models. Specifically, whenever a lead vehicle changes, i.e. $L(i,t)$ changes, then the lead vehicle trajectory $x_{L(i)}$ will have a jump discontinuity, and therefore $h_i$ will have a jump discontinuity as well. \\ 
Given that lane changing always causes discontinuities and multi-regime models may cause discontinuities, one may worry whether or not the objective $F$ will have discontinuities, and whether or not the gradient $dF/dp$ is even guarenteed to exist. We give sufficient conditions for the continuity of $F$ and $dF/dp$ in section \ref{continuous}.

\subsection{Algorithms for Solving the Calibration Problem}
\noindent Eqs. $\eqref{5}$ and $\eqref{6}$ are nonlinear and non-convex optimization problems that must be solved numerically. Current approaches usually rely on gradient-free optimization, with genetic algorithm (GA) and Nelder-Mead simplex algorithm (NM) being the two most commonly applied algorithms \citep{7,8,9,10,11,12,14,19}. These methods have the disadvantage of being slow to converge, and require many objective function evaluations. The objective $F$ is relatively expensive to compute in this context because we first must solve a system of differential equations (i.e. compute the simulated trajectories). As an alternative to directly solving the optimization problems $\eqref{5}$ and $\eqref{6}$, parameters can be estimated directly from data using statistics techniques \citep{2,3,4,51} but this approach is not guaranteed to find local minima. Also, models may have parameters which do not have clear physical interpretations, and therefore cannot be directly estimated. 

Gradient-based optimization algorithms can potentially solve the calibration problem faster than gradient-free approaches. However, even though most optimization algorithms use gradient information, not many gradient-based algorithms have been examined in the calibration literature. To the authors' knowledge, the only gradient-based algorithms that have been considered are simultaneous perturbation stochastic approximation (SPSA) (see \cite{70}), Lindo's Multistart (a commercial solver's global optimization routine, see \citealt{68}), and sequential quadratic programming using filtering (SQP-filter, see \citealt{72}). 
In \cite{60} SQP-filter was found to give better solutions than NM or GA, as well as solve the problem up to 10 times faster; even better results were found when it was additionally combined with gradient-free algorithm DIRECT \citep{62}. \cite{23} did not consider computational speed, but found the Multistart algorithm had a much higher chance of finding the true global minimum compared to GA or NM. 

Even if one only knows how to explicitly evaluate the objective function $F$, it is not necessary to use gradient-free optimization because the gradient (the vector $dF/dp$) can be calculated numerically. To do this, one varies the parameters $p$ one at a time, and recomputes the objective each time a parameter is varied. Then the gradient can be calculated using finite differences. We refer to this as the finite difference approach for computing the gradient, and it requires us to recompute the simulated trajectories of every single vehicle $m$ additional times, where $m$ is the number of parameters. When using a forward euler scheme with stepsize $\Delta t$ for an ODE model, the computation cost for computing the simulated trajectories once is $\mathcal{O}(T(n))$, where $T(n) = \sum_{i=1}^n(T_{i-1} - t_i)/ \Delta t$. So computing the gradient with finite differences has a complexity of $\mathcal{O}(mT(n))$. 
The adjoint method has complexity $\mathcal{O}(T(n))$, and so it is the prefered method for computing the gradient when $m$ is large \citep{22,18}. %Implementing the adjoint method only needs to be considered when $m$ is large.%, which occurs either when the car-following model is complicated, or when a platoon of vehicles is being calibrated (see section \ref{platoonsize}). 

\section{Adjoint Method and Gradient-Based Optimization}\label{adjoint}
\noindent In this paper we are primarily concerned with applying the adjoint method to Eq. $\eqref{5}$ in order to efficiently evaluate the gradient $dF/dp$ for an ODE model. The gradient can then be used with any gradient-based optimization routine. In this section we derive the gradient and compare it to using finite differences; we consider different optimization algorithms in the next section. \\
The adjoint method can also be applied to the calibration problem $\eqref{6}$ where the model includes time delay. It can also be used to calculate the hessian. Both of these extensions follow the same techniques elaborated in this section, and are included in the appendix.  \\
We first consider the calibration of a single vehicle ($n =1$) before dealing with the general case of an arbitrary number of vehicles.
\subsection{Adjoint Method for ODEs}\label{gradient ODE 1}
In our notation, $x_i$, $h_i$, and $p$ are vectors, and $f$ and $F$ are scalars. A derivative of a scalar with respect to a vector is a vector, so for example $\partial f / \partial x_i = [\partial f / \partial x^*_i, \partial f / \partial v_i]$. A derivative of a vector with respect to a vector is a matrix, where the rows correspond to the vector being differentiated, and the columns correspond to the vector being differentiated with respect to. For example, 
\begin{align*} 
\dfrac{\partial x_i}{\partial p} = \bpm \partial x_i^*/\partial p^1 & \partial x_i^* / \partial p^2 & \ldots & \partial x_i^* / \partial p^m \\
\partial v_i / \partial p^1 & \partial v_i / \partial p^2 & \ldots & \partial v_i / \partial p^m \epm
\end{align*}
where $p^j$ represents the $j{\rm th}$ parameter. Additionally, the superscript $^T$ indicates transpose.\\
\noindent Define the augmented objective function, denoted $L$
\[ L = \int_{t_n}^{T_{n-1}} dt \bigg[ f(x_n, \hat x_n, t) + \lambda^T(t)\big( \dot x_n(t) - h_n(x_n(t),x_{L(n)}(t),p_n) \big) \bigg] \tag{9}\label{10}\]
in this equation, $\lambda(t)$ is referred to as the adjoint variable. $\lambda(t)$ has the same dimension as the state variable $x_n(t)$, so it is a length two vector for a second order model. Note that since $\dot x_n - h_n( \cdot) = 0$, we have that $L = F$. These equalities are always true regardless of the choice of $\lambda(t)$. \\
Following the observation that $L = F$, it follows that $dL/dp = dF / dp$, so differentiating Eq. \eqref{10} with respect to the parameters $p$ gives
\[ \dfrac{d}{d p} F = \dfrac{d}{d p} L = \int_{t_n}^{T_{n-1}} dt \bigg[ \dfrac{\partial f}{\partial x_n}\dfrac{\partial x_n}{\partial p} + \lambda^T(t) \bigg( \dfrac{\partial \dot x_n}{\partial p} - \dfrac{\partial h_n}{\partial x_n}\dfrac{\partial x_n}{\partial p} - \dfrac{\partial h_n}{\partial p}\bigg) \bigg] \tag{10}\label{11}\]
where quantities are evaluated at time $t$ if not explicitly stated. Note in this case there are no terms depending on the lead trajectory $x_{L(n)}$ since the platoon consists only of $x_n$ and the lead trajectory is regarded as fixed. \\
Eq. $\eqref{11}$ cannot be computed in the current form because we have no way to calculate ${\partial x_n}/{\partial p}$. The adjoint method works by choosing the the adjoint variable $\lambda(t)$ in a way which allows us to avoid the computation of the unknown ${\partial x_n}/{\partial p}$. When doing this, one will find that the correct choice of $\lambda(t)$ results in solving a similar system to the original constraint, but ``backwards'' (e.g. when the model is an ODE, $\lambda(t)$ will end up being defined as an ODE which is solved backwards in time). The equations which define the adjoint variable are known as the adjoint system. \\
To derive the adjoint system, first apply integration by parts: 
\[  \int_{t_n}^{T_{n-1}} \lambda^T(t)\dfrac{ \partial \dot x_n}{\partial p} dt  = \lambda^T(t) \dfrac{\partial x_n}{\partial p} \bigg|^{T_{n-1}}_{t_n} - \int_{t_n}^{T_{n-1}} \dot \lambda^T(t)  \dfrac{\partial x_n}{\partial p} dt \tag{11}\label{12}\]
The quantity ${\partial x_n(t_{n})}/{\partial p}$ is known because it simply depends on the initial conditions; based on \eqref{5} it is simply zero. We will choose $\lambda(T_{n-1}) = 0$ to eliminate the ${\partial x_n(T_{n-1})}/{\partial p}$ term. This choice becomes the initial conditions for the adjoint system; note that $L = F$ regardless of how the initial conditions for $\lambda$ are chosen. Now combining Eqs. $\eqref{11}$ and $\eqref{12}$: 
\[ \dfrac{d}{dp} F = 
 \int_{t_n}^{T_{n-1}} dt \bigg[ \left( \dfrac{\partial f}{\partial x_n} - \dot \lambda^T(t)- \lambda^T(t) \dfrac{\partial h_n}{\partial x_n} \right) \dfrac{\partial x_n}{\partial p} - \lambda^T(t)\dfrac{\partial h_n}{\partial p}\bigg]  \]
Then to avoid calculating $\partial x_n / \partial p$ we will enforce: 
\[ \dfrac{\partial f}{\partial x_n} - \dot \lambda^T(t)  - \lambda^T(t) \dfrac{\partial  h_n}{\partial x_n} = 0,  \ \ t \in [T_{n-1}, t_n]   \tag{12}\label{13} \]
Eq. $\eqref{13}$ defines a new differential equation  in $\lambda(t)$. Our choice  $\lambda(T_{n-1}) = 0$ is the corresponding initial condition. Then after solving for $\lambda(t)$ we can calculate the desired gradient: 
\[ \dfrac{d}{dp} F = -\int_{t_n}^{T_{n-1}}  \lambda^T(t) \dfrac{ \partial h_n}{\partial p} \ dt \tag{13}\label{14}\] 
 
\subsubsection{N Car Platoon}\label{gradient ODE 2}

\noindent Now we will apply the adjoint method to a platoon of arbitrary size, and sketch a general algorithm for computing the gradient. Define the augmented objective function and proceed as before. Again, we suppress $t$ dependence for clarity. 
\[ L = \sum_{i=1}^{n}\left( \int_{t_i}^{T_{i-1}}f(x_i, \hat x_i) + \lambda_i^T\big( \dot x_i - h_i(x_i, x_{L(i)}, p_i)\big) dt + \int_{T_{i-1}}^{T_i} \lambda_i^T ( \dot x_i - \dot{\hat{x}}_i ) dt\right) \tag{14}\label{15}\]
\begin{align*} 
& \dfrac{d}{dp} F = \dfrac{d}{d p} L  = \sum_{i=1}^{n} \left( \int_{t_i}^{T_{i-1}} \dfrac{\partial f}{\partial x_i}\dfrac{\partial x_i}{\partial p} - \lambda_i^T(t)\dfrac{\partial h_i}{\partial x_i}\dfrac{\partial x_i}{\partial p} - \lambda_i^T(t) \dfrac{\partial h_i}{\partial p}dt \right) \\ 
& +  \sum_{i=1}^{n} \left( \int_{t_{i}}^{T_{i}}\lambda_i^T(t) \dfrac{\partial \dot x_i}{\partial p} - \mathbbm{1}(G(i,t) \neq 0)\lambda^T_{G(i)}(t) \dfrac{\partial h_{G(i)}}{\partial x_i}\dfrac{\partial x_i}{\partial p}dt  \right)
\end{align*}
\noindent Here, $\lambda_n$ corresponds to the adjoint variables for vehicle $n$. Similar to how $L(n,t)$ was defined, $G(n,t)$ is defined as the index of vehicle $n$'s follower at time $t$, so that $x_{G(n)}$ is the following trajectory of vehicle $n$. If vehicle $n$ has no follower at time $t$, \textit{or if the follower is not part of the platoon being calibrated}, let $G(n,t)$ return zero. $\mathbbm{1}$ is the indicator function. When $n>1$, the downstream boundary conditions must be put into the augmented objective function, because a vehicle $i$ may affect its follower in the time $[T_{i-1}, T_i]$, even though it only contributes to the loss function during $[t_i, T_{i-1}]$. \\
Apply integration by parts, take the initial condition as $\lambda_i(T_{i}) = 0$, and gather terms with $\partial x_i / \partial p$ to form the adjoint system: 
\begin{align*} 
& - \dot \lambda_i^T(t) + \mathbbm{1}(t \leq T_{i-1}) \left( \dfrac{\partial f}{\partial x_i} - \lambda_i^T(t)\dfrac{\partial h_i}{\partial x_i} \right) - \mathbbm{1}(G(i,t) \neq 0) \lambda_{G(i)}^T(t) \dfrac{\partial h_{G(i)}}{\partial x_i} = 0, \quad t \in [T_{i}, t_i],\ \forall i \tag{15}\label{16}
\end{align*}
Comparing this adjoint system to the one for a single vehicle \eqref{13}, the difference here is that there is an extra contribution which occurs from the coupling between vehicles. This coupling term occurs only when a vehicle acts as a leader for another vehicle in the same platoon.  
After solving the adjoint system, 
\begin{align*} 
& \dfrac{d}{dp }F  = \sum_{i=1}^n \left(  - \int_{t_i}^{T_{i-1}} \lambda_i^T(t)\dfrac{\partial h_i}{\partial p}dt  \right) \tag{16}\label{18}
\end{align*}
All the partial derivatives in Eqs. $\eqref{16}, \eqref{18}$ are computed exactly for a given loss function and car-following model. Solving for $x_i(t), \lambda_i(t), \text{ and } d F / dp$ is done numerically with an appropriate time discretization. The same discretization should be used for all quantities, so if one uses a forward euler scheme for $x_i$, the same forward euler scheme should be used for $\lambda_i$ and the integral in $F$ should use a left Riemann sum. Note as well that for a multi-regime model, one should keep the regime for each timestep in memory, because otherwise the switching condition $g_{i,n}$ will have to be evaluated at each timestep when computing the adjoint variables. \\
We define a \textbf{general algorithm for computing $F$, $\dfrac{d}{dp} F$:} \\
\textbf{Inputs: }measured trajectory data $\hat x_1(t), \ldots, \hat x_n(t)$, any necessary lead vehicle trajectories $x_{L(i)}(t)$ for $i \in [1, \ldots, n], \ L(i, t) \notin [1, \ldots, n]$. Models $h_1, \ldots, h_n$ with parameters $p_1, \ldots, p_n$.
\begin{enumerate}
\item For each vehicle $i \in [1, \ldots, n]$:
\begin{enumerate}
\item Compute simulated trajectory $x_i(t)$, $t \in [t_i, T_{i-1}]$ by solving $\dot x_i(t) = h_i( x_i, x_{L(i)}, p_i)$ with intial condition $x_i(t_i) = \hat x_i(t_i)$. 
\item Compute $x_i(t)$, $t \in [T_{i-1}, T_i]$ using downstream boundary conditions Eq. \eqref{7}
\item Keep in memory: $x_i(t)$, $G(i,t)$, and the regime of $h_i$ at each time
\end{enumerate}
\item Compute objective $F = \sum_i^n\int_{t_i}^{T_{i-1}} f( x_i, \hat x_i, t) dt$
\item For each adjoint variable $i \in [n, \ldots, 1]:$ compute $\lambda_i(t)$, $t \in [T_i, t_i]$ by solving adjoint system Eq. $\eqref{16}$ with initial condition $\lambda_i(T_{i}) = 0$.
\item Compute gradient $dF / dp = -\sum_{i}^n\int_{t_i}^{T_{i-1}}\lambda_i^T\partial h_i / \partial p \ dt$
\end{enumerate}
\textbf{Outputs: } $F$, $d F / dp$
\subsubsection{Sufficient Conditions for a Continuous Objective and Gradient} \label{continuous}
\begin{theorem}
The following are sufficient conditions for $F$ and $dF / dp$ to be continuous:
\begin{enumerate}
\item the loss function $f(x_i, \hat x_i)$ and its partial derivative $\partial f/ \partial x_i$ are piecewise continuous on $[t_i, T_{i-1}]$ for all $i$
\item the car following model $h_i ( x_i, x_{L(i)}, p_i )$ and its partial derivatives $\partial h_i / \partial x_i$, $\ \partial h_i / \partial p$, and $\partial h_i / \partial x_{L(i)}$ are piecewise continuous on $[t_i, T_{i-1}]$ for all $i$
\item $\partial h_i / \partial x_i$ must be able to be bounded by a constant 
\item for a multi-regime model, the Eq. \ref{condition} must be satisfied
\end{enumerate}
where it is additionally assumed that $x_{L(i)}(t), L(i,t) \notin [1, \ldots, n]$ (any lead vehicle trajectories which are required but not simulated) are piecewise continuous and $\hat x_i(t), i \in [1, \ldots, n]$ are continuous.
\end{theorem}
\begin{corollary} 
If in addition to $F$ and $dF / dp$ being continuous, $\partial f / \partial x_i, \ \partial h_i / \partial x_{L(i)}, \text{and} \ \partial h_i / \partial p$ are bounded, then $F$ is Lipschitz continuous. 
\end{corollary}
We conclude that a multi-regime model need not be continuous when switching between regimes, and lane changing, which will always cause jump discontinuities in $h_i(x_i, x_{L(i)}, p_i)$, is allowed. Note that by piecewise continuity we mean a model (or loss function) can switch regimes only a finite number of times in a finite time interval (and each regime is continuous); additionally, the time interval of each regime must be continuous with respect to the model parameters. Refer to the the full proof in the appendix for further details. We also show in the appendix that the objective is Lipschitz continuous for the OVM, which was the model used for the numerical experiments in this paper.

\subsection{Implementation of Adjoint Method and Calibration Problem} \label{implementation}
To test the adjoint method and optimization algorithms, the calibration problem Eq. \eqref{5} was implemented in python for the optimal velocity model. The equations below show the functional form of the model as originally formulated in \cite{51} and \cite{52}. 
\bg{ \ddot x_n(t) = c_4(V(s) - \dot x_n(t))  
\\ V(s) = c_1[ \tanh(c_2 s - c_3 - c_5) - \tanh(-c_3)] }
$c_1$ through $c_5$ are the 5 parameters of the model. $s$ is the space headway, defined as $s = x_{L(i)} - x_i - l_{L(i)}$, where $l_i$ is the length of vehicle $i$. All vehicles have their own unique parameters, so the total number of parameters in the optimization problem is 5 times the number of vehicles. 

The optimal velocity model and its adjoint system were both discretized with a forward euler scheme. The data used was the reconstructed NGSim data \citep{29}, so a time step of .1 seconds was used to be consistent with this data. The loss function $f$ used was square error, which was discretized with a left Riemann sum. 
\begin{align*} 
f(x_i, \hat x_i, t) = (x_i^*(t) - \hat x_i^*(t))^2 \\
F = \sum_{i=1}^n \sum_{t = t_i}^{T_{i-1}} f(x_i, \hat x_i, t)
\end{align*}
which is equivalent to minimizing the root mean square error (RMSE)
\begin{align*} 
\text{RMSE} = \sqrt{\dfrac{F}{\sum_{i = 1}^n T_{i-1}- t_i}}
\end{align*}
Since only a car following model is being calibrated, any lane changes in the measurements will be the same lane changes in the simulation. This also means that the leader/follower relationships in the simulation are the same as they are in the data. This allows trajectories to be calibrated at the level of individual vehicles, since it does not make sense to directly compare the simulated and measured trajectories of a vehicle if the simulation and measurements have different lead vehicles. 

\subsection{Adjoint method compared to other methods for calculating the gradient}

Many optimization algorithms will automatically use finite differences to compute the gradient if an explicit gradient function is not given. Simultaneous perturbation is another option when using an optimization algorithm like the one outlined in \cite{70}. Then, assuming that one has chosen to use a gradient-based algorithm for calibration, it is of practical interest to investigate how the adjoint method compares to both of these methods. Specifically, how does the speed and accuracy of these three methods compare in practice? 
\subsubsection{Comparison of Speed} \label{adjointspeedsection}
The computational cost of an objective function evaluation is $\mathcal{O}(T(n))$, where $T(n)$ is the number of simulated timesteps, added over all $n$ vehicles. The theoretical computational complexities for the adjoint method is $\mathcal{O}(T(n))$, while forward differences has $\mathcal{O}(mT(n))$, where $m$ is the number of parameters. The simultaneous perturbation method has $\mathcal{O}(kT(n))$ complexity, where the final gradient is the average of $k$ estimates of the gradient. Note that these computational complexities are only for the gradient calculation and do not consider the complexity of whatever optimization algorithm the gradient is used with.

\begin{figure}[h] 
\centering 
\includegraphics[ width=\textwidth]{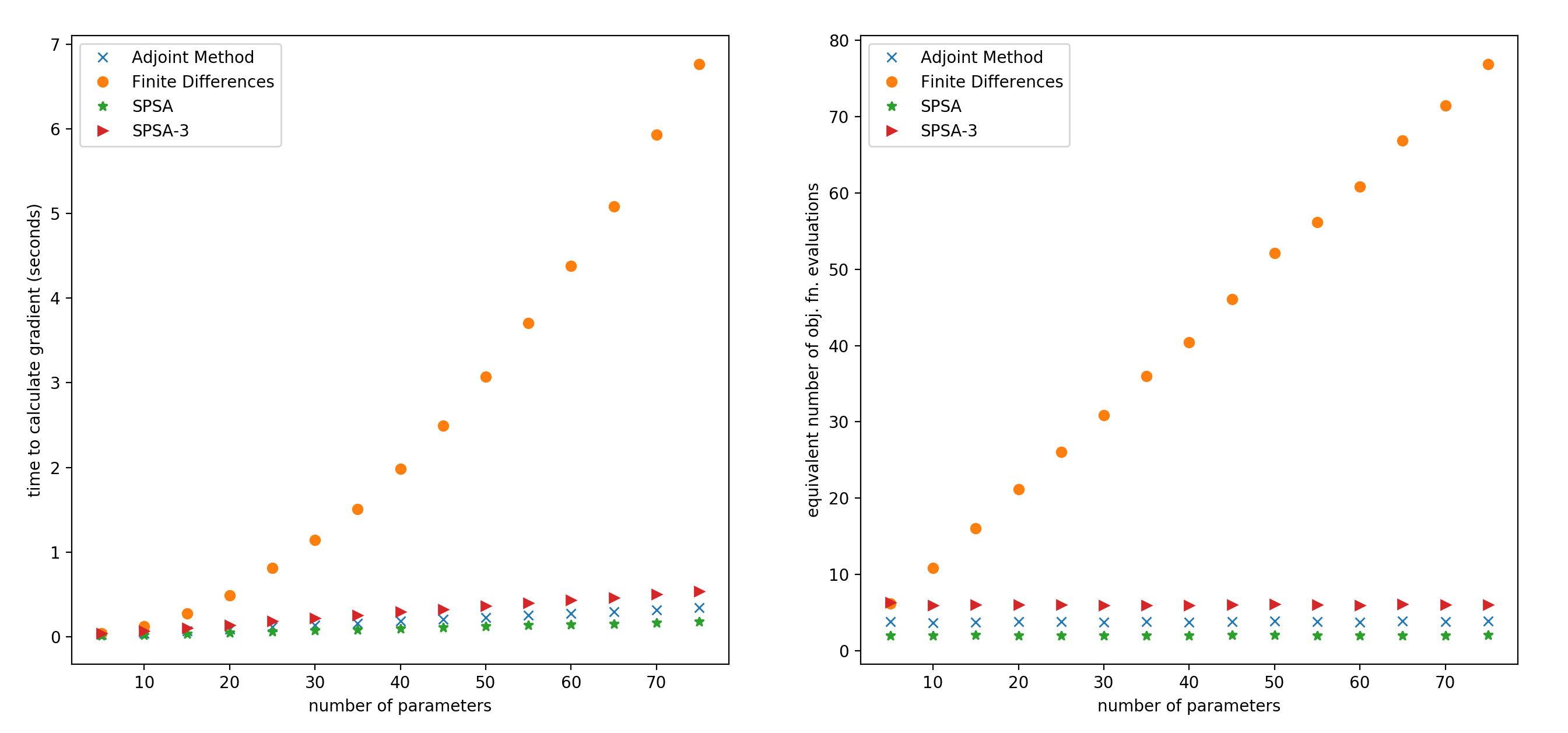} 
\caption{Time for a single evaluation of the gradient for a progressively larger calibration problem. 
The objective varied from the loss from a single vehicle (5 parameters) to 15 vehicles (75 parameters).}  \label{adjointspeed}
\end{figure}

To test these theoretical results in practice, the gradient of the calibration problem was calculated and the platoon size was varied. Since the OVM was used, the total number of parameters is 5 times the number of vehicles. We considered a platoon having between 1 and 15 vehicles. The results are shown in the figure \ref{adjointspeed}. 

\noindent Figure \ref{adjointspeed}'s left panel shows the total CPU time, measured in seconds, needed to calculate the gradient of the calibration problem using the three different methods. The simultaneous perturbation was done with both 1 and 3 trials, denoted SPSA and SPSA-3. The right panel shows the time needed to calculate the gradient divided by the time needed to calculate the objective. To minimize the randomness in the CPU times, each data point was repeated 10 times and the results were averaged. Some variation is still present because some vehicles are observed for shorter times. The same 15 vehicles were used in the experiment and always added in the same order. It should also be stated that for either a finite differencing scheme or for the adjoint method, one needs to first compute the objective before computing the gradient. Therefore, it should be implicitly understood that ``time to calculate the gradient", really refers to ``time to calculate both the objective and gradient" for those two methods. 
\noindent This shows that using a finite difference method quickly becomes intractable when the number of parameters increases. The time increases quadratically in the left panel because both $m$ and $T(n)$ are increasing. In the right panel, when the time needed to calculate the gradient is divided by the time needed to calculate the objective, one sees the complexity growing as a linear function of $m$. In that figure, we also see that the complexities for the adjoint method and simultaneous perturbation do not depend on $m$. 

\noindent The table \ref{speedtable} shows the cpu times for 5-15 parameters. 
One can observe that this implementation of the adjoint method for OVM requires the equivalent of 4 objective function solves, whereas forward differences requires $m+1$ (where there are $m$ parameters).

\begin{table}[h]
\caption{Table of speeds for adjoint method compared to finite differences for computing the gradient. Relative time refers to the equivalent number of objective function evaluations.  } \label{speedtable}
\begin{tabular}{|l|l|l|l|l|l|}
\hline
Parameters & Obj. time (s) & Adjoint time (s) & Adjoint relative time & Finite time (s) & Finite relative time \\ \hline
5                       & .0062         & .0250            & 4.03          & .0390           & 6.3         \\ \hline
10                      & .0109         & .0438            & 4.02          & .1250           & 11.5        \\ \hline
15                      & .0172         & .0686            & 3.99          & .2811           & 16.3        \\ \hline
\end{tabular}
\end{table}

\subsubsection{Comparison of Accuracy} \label{adjointacc}

For the calibration problem, there is no way to calculate the gradient exactly because there is no closed form solution for the car following model. To evaluate the accuracy of the adjoint method and simultaneous perturbation, we will therefore treat the result from forward differences as the true solution. In the experiment, points were evenly sampled between an initial starting guess and the global minimum of the calibration problem for a single vehicle. At each of these points in the parameter space, the relative error was calculated
$$\text{relative error} = \dfrac{||\nabla_{fd}F - \nabla F||_2}{||\nabla_{fd}F||_2}$$ 
where $\nabla_{fd}F$ is the gradient calculated with forward differences, $\nabla F$ is the gradient calculated with the adjoint method or simultaneous perturbation, and $||\cdot ||_2$ denotes the 2 norm. The results of the experiment are shown below.

\begin{figure}[h] 
\centering 
\includegraphics[ width=\textwidth]{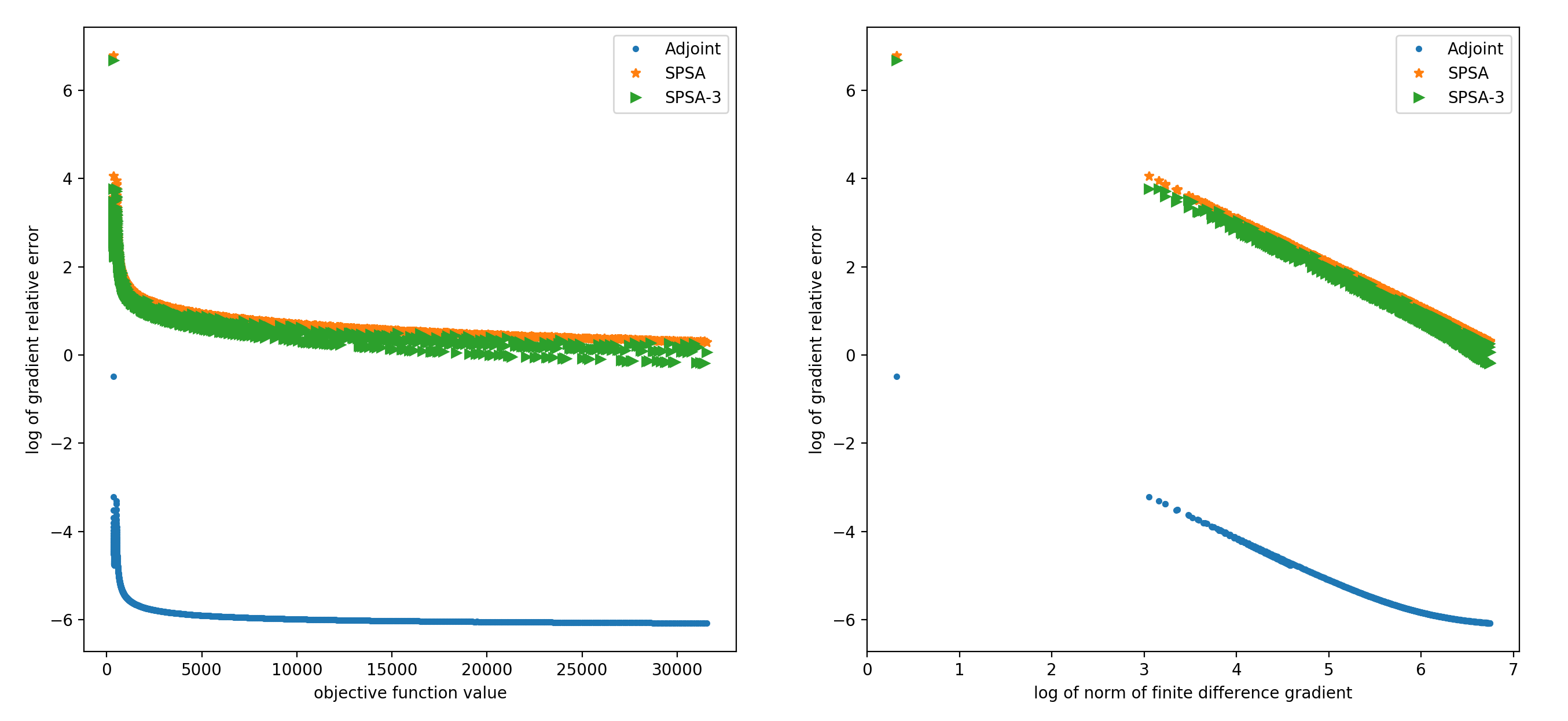} 
\caption{Compares the accuracy of the gradient for the adjoint method and simultaneous perturbation. The finite differences gradient is treated as the exact gradient. } 
\end{figure}

\noindent For the adjoint method, the relative error is usually small, on the order of $10^{-6}$. When the parameters start to become close to a local minimum, the relative error starts to rapidly increase to the order of $10^{-4}-10^{-3}$. When at the optimum, the relative error becomes as high as .3. The reason why the relative error increases is because as the parameters approach the minimum, the norm of the gradient decreases. The right panel shows the relationship between the log of the norm of the gradient calculated with finite differences, and the log of the relative error. One can see that they are inversely proportional to each other. This makes theoretical sense: we expect the residual $\nabla_{fd}F-\nabla F$ to always be on approximately the same order of magnitude since it depends mainly on truncation error which does not change with the parameters. Then, it follows that the relative error is dominated by the $||\nabla_{fd}F||$ term in the denominator, which is what we observe. 

\noindent The relative error for simultaneous perturbation follows the same shape as the adjoint method, but is always about 6 orders of magnitude larger. Of course, you would not use simultaneous perturbation as a substitute for the finite difference gradient, since it is used with its own optimization algorithm. Nonetheless, this shows that although the two methods (adjoint and simultaneous perturbation) both have a flat computational cost with the number of parameters, they have different uses. The reason why the the error for simultaneous perturbation is quite high in this context is because the gradient is scaled very poorly for the OVM model, meaning the parameters have very different sensitivities. This poor scaling is a common feature for car following models. 

\section{Benchmarking different Algorithms and the Adjoint method} \label{algos}
 The ideal optimization algorithm would be one that can not only solve the problem quickly, but also one that is able to avoid bad solutions corresponding to local extrema. Because existing literature has focused on gradient-free optimization algorithms, not much is known about using gradient/Hessian-based optimization to solve the calibration problem. The only works the authors know of in the car following calibration literature that consider gradient based optimization are \cite{60} and \cite{23}. Both those works found the gradient based algorithms to work best overall. This result is consistent with other engineering applications; for example \cite{64} and \cite{65} deal with aerodynamic shape optimization and material science respectively, and report that gradient-based optimization can offer significant speed increases.

To establish the benefits of gradient based optimization and the adjoint method, we consider 2 gradient-free and 5 gradient-based optimization algorithms. The different algorithms, their specific implementations, and abbreviations used throughout the rest of the paper, are detailed below. 
\begin{itemize}
\item Genetic Algorithm (GA) -  an initial population of trial solutions is randomly selected throughout the parameter space. In each iteration, the population is altered through stochastic processes referred to as ``mutation" and ``crossover", and then updated in a selection step. An implementation following \cite{73} is part of the scipy package in python. 
\item Nelder-Mead Simplex (NM) - a given initial guess is used to define a collection of points (a simplex). In each iteration, the corners of the simplex are updated according to deterministic rules. Implemented in scipy following \cite{74}. NM works on unconstrained problems, so a large penalty term was added to the objective if the parameters went outside the bounds.
\item Limited memory BFGS for bound constraints (BFGS) - at every iteration, gradient information from the previous iterates is used in the BFGS formula to form an approximation to the hessian. A quadratic approximation to the objective is formed, and the generalized cauchy point of this approximation is then found, which defines a set of active bound constraints. A minimization is then performed over the variables with inactive constraints, while the variables with active constraints are held fixed. The solution to this last minimization finally defines the search direction for the algorithm, and a line search is performed to determine the step length. The algorithm, referred to as l-bfgs-b, is due to \cite{75} and \cite{76} and wrapped as part of the scipy package. 
\item Truncated Newton Conjugate (TNC) - at every iteration, instead of explicitly computing the Newton direction $-\nabla^2 F^{-1} \nabla F$, the conjugate gradient method is applied to the newton system $\nabla^2 F d = -\nabla F$ (where $d$ is the search direction to be determined). Conjugate gradient is itself an iterative algorithm, and it is terminated early (truncated) to give an approximate solution to the newton system which is also guaranteed to define a direction of descent. A line search is then performed to define the step length. The conjugate gradient algorithm only needs hessian-vector products which can be computed efficiently using finite differences, so the full hessian never needs to be calculated. An implementation due to \cite{78} is wrapped as part of scipy. That implementation uses a conjugate gradient preconditioner based on the BFGS update, and the search direction is projected to enforce the bound constraints.
\item Simultaneous Perturbation Stochastic Approximation (SPSA) - At each iteration, the gradient is approximated using simultaneous perturbation. Then the algorithm moves in the direction of the negative gradient with a fixed step length, so no line search is performed. The size of the step length is gradually reduced in subsequent iterations. Implemented in python following \cite{70}; the step sizes and number of iterations were manually tuned by hand. 
\item Gradient Descent (GD) - At every iteration, the negative gradient is computed and scaled according to the spectral step length (also known as the Barzilei Borwein method). The scaled gradient is then projected onto the bound constraints to define the search direction, and a nonmonotone backtracking linesearch is used to define the step length. Implemented according to \cite{77} in python. The algorithm was tested with different line searches (backtracking, weak/strong wolfe, watchdog, see \citealt{71}) and the nonmonotone backtracking gave the best performance. 
\item Sequential quadratic programming (SQP) - At every iteration the hessian is computed and the search direction is defined by the newton direction projected onto the bound constraints. This algorithm explicitly computes the hessian instead of approximating it. Because the hessian may fail to be positive definite, the search direction is safeguarded so that if the newton direction fails to define a direction of descent, the algorithm instead searches in the direction of the negative gradient. A small regularization is also added to the hessian so that it can always be inverted. The algorithm is a simple python implementation of sequential quadratic programming based on \cite{71}. We again found the nonmonotone backtracking linesearch to be most effective. 
\end{itemize}

In summary, out of all the algorithms tested: NM and GA are gradient-free; SPSA and GA are stochastic; BFGS and TNC are quasi-newton; GD and SPSA do not use hessian information; SQP explicitly computes the hessian; GA is the only global algorithm.

For the TNC, BFGS, GD, and SQP algorithms, we considered two separate modifications for each algorithm. First, either the finite difference method or the adjoint method can be used to compute the gradient, and we denote this by either ``Fin" or "Adj" respectively. Second we considered starting from multiple initial guesses. Multiple initial guesses are specified, and after each calibration, if the RMSE was above a specified threshold, the next initial guess would be used. The same 3 initial guesses were used in the same order for all algorithms requiring an initial guess. So with a threshold of 0, 3 guesses are always used, and with an arbitrarily large threshold, only a single guess will be used. The thresholds of 0, 7.5 and $\infty$ were used, and denoted in reference to the algorithm. The initial guesses used are based off of the calibrated parameters found from \cite{51}.

\subsection{Evaluating Algorithm Performance}
As detailed in section \ref{implementation}, we tested the calibration problem on the OVM car following model using squared error in distance as the loss function (equivalent to minimizing the RMSE). In this experiment, for each algorithm, every vehicle in the reconstructed NGSim dataset was calibrated using the true measurements of its lead trajectory. In total, there were 7 unique algorithms and 23 total algorithms tested when including the variants described above. 

The calibration results are evaluated primarily based on three different measures: the average calibration time for a single vehicle, the \% of the time the algorithm found the global minimum, and the average RMSE of a calibrated vehicle. Since there is no way to know the actual global minimum for the calibration problem, we regard the global minimum for a vehicle as being the best result out of all the algorithms, and we regard an algorithm as having found the global minimum if its RMSE is within 1/12 ft (1 inch) of this best result. 

\begin{figure}[h] 
\centering 
\includegraphics[ width=\textwidth]{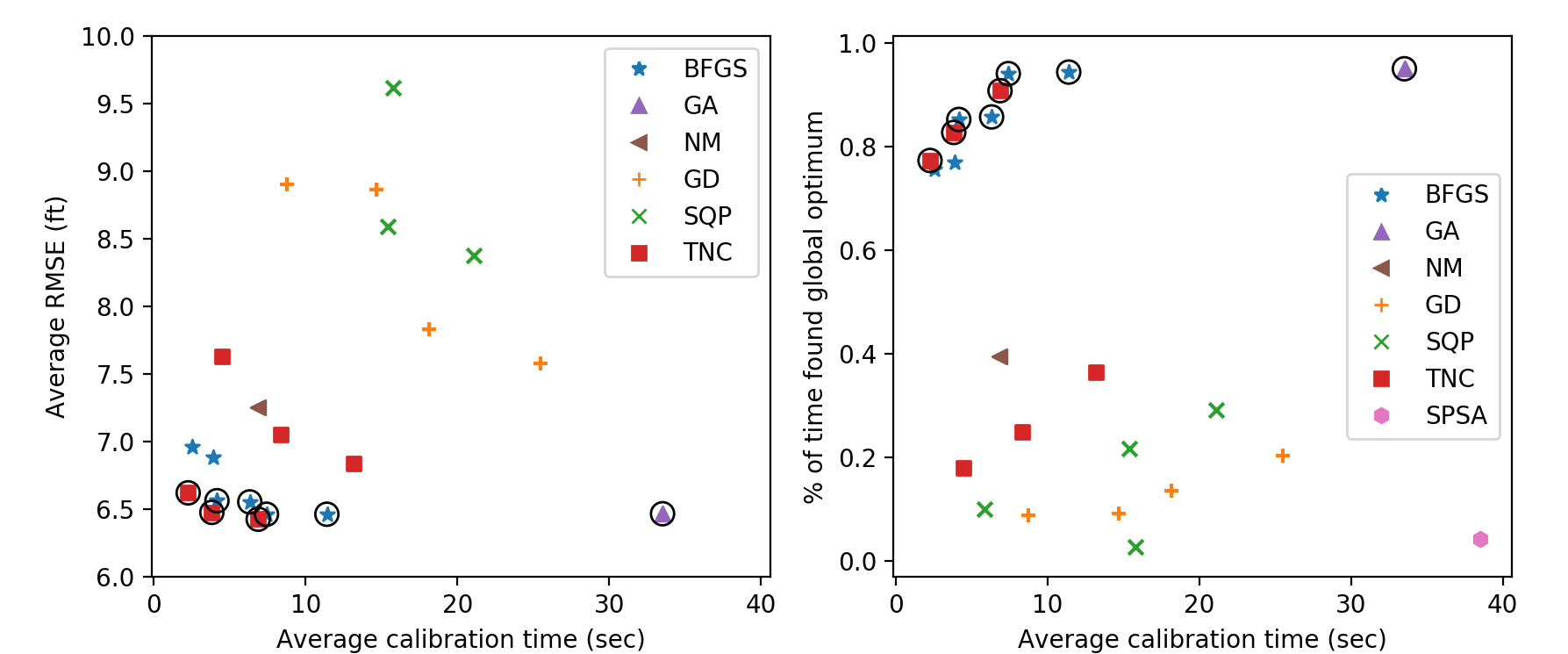} 
\caption{All algorithms plotted by the three metrics time, RMSE, and \% found the global optimum. Algorithms on the pareto front with respet to these three metrics circled in black.}  \label{paretofig}
\end{figure}

Based on these three metrics, all algorithms on the pareto front were identified and are shown in the table \ref{table2}. 
In figure \ref{paretofig} every algorithm is plotted in the RMSE - time and \% found global optimum - time plane. The algorithms on the pareto front are circled in black. 

\begin{table}[h]
\caption{Algorithms on the Pareto front. All algorithms tested shown in appendix.} \label{table2}
\begin{tabular}{|l|l|l|l|l|l|l|l|l|}
\hline
Algorithm & \begin{tabular}{@{}c@{}}\% found  \\global opt\end{tabular} &  \begin{tabular}{@{}c@{}} Avg.  \\ RMSE (ft) \end{tabular} &  \begin{tabular}{@{}c@{}} Avg. \\  Time (s)\end{tabular} & \begin{tabular}{@{}c@{}} Avg. RMSE / \% \\  over global opt\end{tabular} &\begin{tabular}{@{}c@{}}Initial  \\ Guesses\end{tabular} & \# Obj. Evals & \# Grad. Evals\\ \hline
Adj BFGS-0 & 94.0  & 6.46& 7.42 & .10 / 2.0\% &3 &307.1 & 307.1 \\ \hline
Adj BFGS-7.5 &85.2 &6.56 &4.16   & .20/6.8\%& 1.67 & 165.0&  165.0\\ \hline
GA & 95.0& 6.47 &33.5  & .10/2.2\% & -  & 5391 & 0\\ \hline
Fin BFGS-0& 94.3 & 6.46 & 11.4  & .10/2.2\% &3 & 311.5 &311.5  \\ \hline
Fin BFGS-7.5&85.7 & 6.55 & 6.32   & .19/6.3\%&1.66 & 164.8& 164.8\\ \hline
Adj TNC-0& 90.7  & 6.43 & 6.87 & .05/2.4\% &3 & 285.7 & 285.7 \\ \hline
Adj TNC-7.5& 82.7  & 6.48 & 3.82  & .11/5.0\%& 1.63 & 152.3 &152.3  \\ \hline
Adj TNC-$\infty$& 77.3 &  6.62& 2.26 & .25/7.7\%& 1 & 94.1 & 94.1 \\ \hline
\end{tabular}
\end{table}

The only three algorithms on the pareto front were the GA, TNC, and BFGS. GA gave the best \% of finding the global optimum. TNC with the adjoint method and three guesses gave the best overall RMSE, and was the fastest algorithm overall when used with the adjoint method and a single guess. BFGS was slightly slower than TNC and has a slightly higher RMSE, but it has a higher chance of finding the global optimum. 

\begin{figure}[h] 
\centering 
\includegraphics[ width=\textwidth]{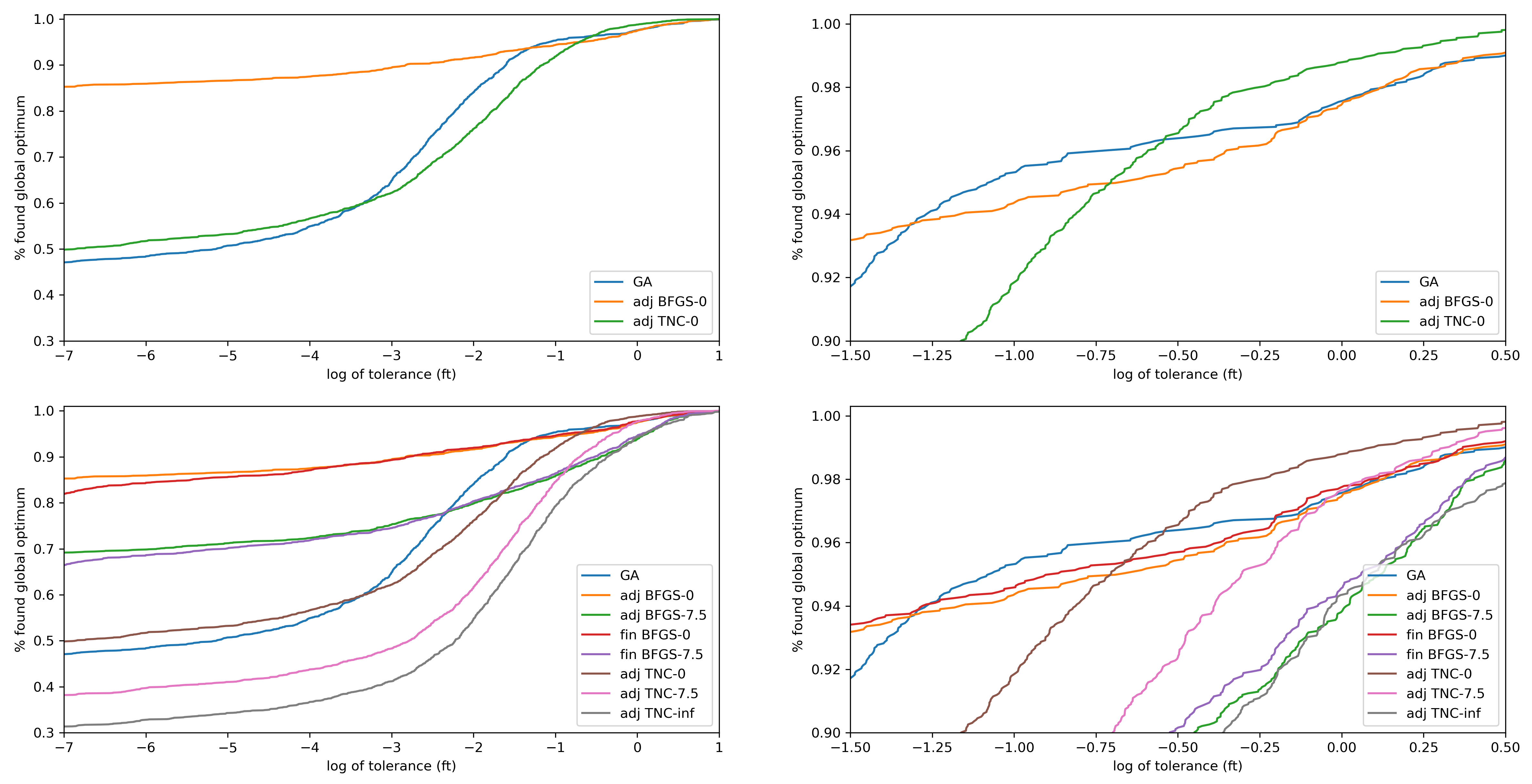} 
\caption{Shows the distribution of RMSE over the global optimum for GA, Adj TNC-0, Adj BFGS-0 in top two panels. Bottom two panels show the distributions over all algorithms on the pareto front (with a tolerance of 1/12 used for obtaining the pareto front). Right panels are zoomed in windows of left panels.} \label{rmsedist}
\end{figure}

The conclusion of what algorithms are on the pareto front depends strongly on how ``finding the global optimum" is defined. Here, we treat an algorithm as having found the global optimum if it gives a result within an inch (1/12) of the best RMSE. Let us refer to this value of 1/12 as the ``tolerance" for finding the global optimum. If a smaller tolerance is used, then GA will no longer be on the pareto front. If a larger tolerance is used, GA and BFGS will no longer be on the pareto front. 
For example, with a tolerance of 0, the pareto front consists of only the adj TNC and adj BFGS variants, and adj BFGS-0 gives the highest chance of finding the global optimum. With a tolerance of 1/2, the adj TNC variants are the only algorithms on the pareto front. The effects of changing the tolerance are shown in the table \ref{table3}. Adj TNC will always be on the pareto front regardless of the tolerance because it gives the best average RMSE for any given speed. No algorithms other than adj TNC, BFGS, and GA are ever on the pareto front. 
\begin{table}[h]
\centering 
\caption{Algorithms on the Pareto front depend on how ``finding the global optimum" is defined (without that metric, TNC with the adjoint method would be the only algorithm on the pareto front). 
Set notation indicates the strategies used for initial guesses.} \label{table3}
\begin{tabular}{|l|l|}
\hline
 \begin{tabular}{@{}c@{}}Tolerance  for  \\global opt (ft) \end{tabular}  &  Algorithms on the pareto front\\ \hline
0 & Adj TNC-\{0, 7.5, Inf\}, Adj BFGS-\{0, 7.5, Inf\} \\ \hline
1/24 & Adj TNC-\{0, 7.5, Inf\}, Adj BFGS-\{0, 7.5, Inf\}, Fin BFGS-\{0, 7.5\} \\ \hline
1/12 & Adj TNC-\{0, 7.5, Inf\}, Adj BFGS-\{0, 7.5\}, Fin BFGS-\{0, 7.5\}, GA \\ \hline
1/4 & Adj TNC-\{0, 7.5, Inf\}, GA \\ \hline
1/2 & Adj TNC-\{0, 7.5, Inf\}\\ \hline
\end{tabular}
\end{table}
The behavior shown in table \ref{table3} is explained by figure \ref{rmsedist}. That figure shows how the \% found global optimum metric changes with tolerance. For example, a tolerance of 1/12 corresponds to -1.08 in a log scale, and the top right panel shows that at that point, the GA has the highest chance of finding the global optimum. The same panel shows that if the tolerance increases, adj TNC-0 will have the highest chance, and that adj BFGS-0 will have the highest chance if the tolerance decreases. 

BFGS typically found the best solution out of all the algorithms tested, but its distribution of RMSE has a heavier tail in the sense that there were also some vehicles for which BFGS couldn't find a good solution. TNC is most consistent because its distribution shows less of a heavy tail (it is the first to reach 100\% found global optimum as the tolerance increases). The GA has a tail similar to BFGS and performs well when only a moderate tolerance is needed. 

The appendix includes a plot that shows the distributions of RMSE (on a non-log scale), and the differences between the three algorithms are essentially indistinguishable at that resolution. 

Overall, these differences between the three algorithms (GA, adj TNC-0, adj BFGS-0) are relatively minor, with all three giving essentially the same average RMSE (GA had a slightly worse RMSE than the other two). It's not clear to the authors whether the subtle differences between them would cause any differences in practice. However, the BFGS and TNC algorithms both give a significant speed increase; the adjoint method with BFGS or TNC is about 5 times faster than the GA on average. Even when using finite differences instead of the adjoint method, BFGS is still about 3 times faster than the GA (finite differences with TNC did not work well). A much larger speed increase can be realized with only a small trade off in accuracy. Adj TNC-Inf had an average RMSE of 6.62 compared to 6.47 for the GA, but is 15 times faster. Adj TNC-7.5 had an average RMSE of 6.48 while being 9 times faster than the GA.

The comparison between the different strategies for initial guesses (0, 7.5, Inf) shows the trade off between speed and accuracy when starting a local search based algorithm multiple times. For car following models, it seems that even a suboptimal local minimum still gives a reasonable accuracy. The difference in accuracy made by starting a algorithm multiple times is smaller than the difference from using another algorithm, so regardless of the initial guess, the local search seems to converge to similar parameters. We also saw that a good local search algorithm (in this case, the quasi-newton algorithms) can give results equivalent to a global search algorithm (the GA) when a small number of initial guesses are used. Of course, each initial guess used greatly increases the time needed for the algorithm, so a hybrid strategy similar to the one proposed in \cite{60} can be considered if a large number of initial guesses are needed to achieve good performance. 

Comparing the results of using the adjoint method to finite differences, the differences between the two methods depends heavily on the algorithm used. For BFGS, using finite differences instead of the adjoint method gives essentially the same result, in the same number of objective/gradient evaluations, and the difference is speeds is consistent with the speed increase found in section \ref{adjointspeed}. For TNC, the algorithm does not perform well when finite differences is used. 

All the attention so far has been concentrated on the TNC, BFGS, and GA algorithms since those gave good results. None of the other algorithms performed well. NM is designed for unconstrained problems, so it is unsurprising that it did not perform as well as an algorithm designed specifically to deal with box constraints (recall that here the box bounds were enforced by adding a penalty term). GD suffered from the problem widely reported in the literature where it takes a very large number of iterations to fully converge. The majority of the time, the algorithm terminates due to exceeding the maximum number of evaluations allowed. For the SQP (recall in this case we explicitly computed the Hessian), we found that the newton direction often fails to define a direction of descent, and the algorithm simply searches in the direction of the gradient instead. Then, the algorithm spends a lot of computational time computing the Hessian but ends up not using it for anything, so explicitly computing the hessian seems to not work poorly for car following calibration; BFGS updating to estimate the Hessian, or a truncated newton method for estimating the newton direction should be used instead. SPSA performed worst out of all algorithms tested because of the vastly different sensitivities for car following models, as well as the rapidly changing magnitude of the gradient. A variant of SPSA such as c-SPSA \citep{cspsa} or W-SPSA \citep{wspsa} may have overcome these issues. 

The analysis in this section has shown the value of gradient based algorithms and the adjoint method; compared to gradient free methods, they can give the same overall performance while offering a significant speed increase. We found that for the calibration of a single vehicle, the truncated newton (TNC) algorithm with the adjoint method gives a slightly more accurate calibration than a genetic algorithm, while also being 5 times faster. TNC is 15 times faster than the genetic algorithm if $\approx$ 2\%  decrease in accuracy is acceptable (when only a single initial guess is used). Actually, as we will see in the next subsection, when the calibration problem becomes larger, the benefit of gradient based algorithms and the adjoint method becomes even larger.

\subsection{Calibration of larger platoons}\label{platoonsize}
\noindent Define platoon size as the number of vehicles calibrated at the same time ($n$ in Eq. \eqref{5}). The simplest case is when $n=1$: a single vehicle is calibrated at a time (this was the strategy used in the above section). Taking $n=1$ is how most of the literature treats car following calibration. Moreover, trajectories are typically calibrated to the measured, as opposed to simulated, lead trajectory. To the authors' knowledge, \cite{19} is the only paper which has taken a platoon size of larger than 1 ($n=2$ in that paper) and \cite{recurrentcf} is the only paper which considers the issue of predicting on the simulated vs. measured lead trajectory. In actual traffic simulations, all vehicles are simulated at the same time, and measured trajectories aren't assumed to be available. In that case, errors from one vehicle can accumulate in the following vehicles and produce unrealistic effects. This motivates the question of calibrating platoons of vehicles, where each vehicle uses the simulated lead trajectory.

In this section, we are interested in two questions regarding the calibration for a platoon size of $n>1$. First, for the different methods considered so far (gradient free, gradient based with finite differences, adjoint method) how does the time needed to solve the calibration problem scale with the number of parameters. Second, what is the benefit (in terms of RMSE) from considering a larger platoon of vehicles.

A platoon of 100 vehicles was formed from the NGSim data. The platoon was formed by randomly selecting the first vehicle to be calibrated, and then repeatedly adding vehicles which have their leaders in the platoon. We considered calibrating these 100 vehicles with a platoon size $n = 1, 2, \ldots 10$. When the platoon size doesn't divide 100, the last platoon used will be the remainder (e.g. for $n = 3$, the last platoon is of size 1). Here the calibration is done sequentially, so each vehicle is calibrated to the simulated trajectory of its leader. 

\begin{figure}[h] 
\centering 
\includegraphics[ width=.65\textwidth]{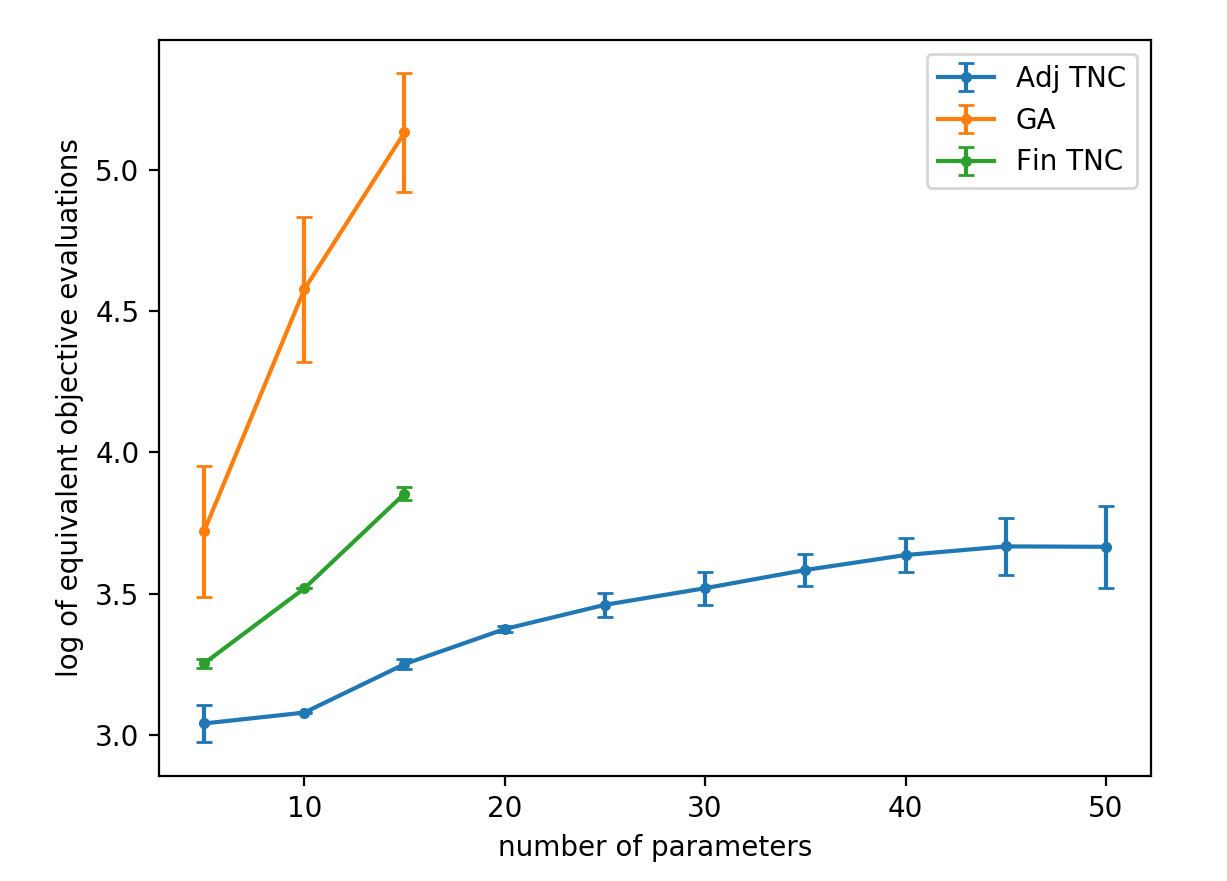}  
\caption{Log (base 10) of equivalent number of objective evaluations required to solve the calibration problem for an increasing platoon size ($n)$. The number of parameters in each platoon is $5n$. The brackets show the standard deviation for the data points. The work required to solve the calibration problem increases at a much faster rate when using gradient free optimization or finite differences compared to the adjoint method.  } \label{speed2}
\end{figure}

To compare how different algorithms scale, we performed calibration using the GA, Adj TNC, and Fin TNC algorithms, as shown in figure \ref{speed2}. For each platoon of a specific size, the number of equivalent objective evaluations is recorded. Gradient evaluations are converted into objective evaluations following the results of \ref{adjointspeedsection} (adjoint method can compute the gradient and objective in the equivalent of 4 objective evaluations, finite differences computes the gradient and objective in the equivalent of $m+1$ objective evaluations, where $m$ is the number of parameters). The results are shown in figure \ref{speed2}.

\begin{table}[h]
\centering
\caption{Compares the number of equivalent objective evaluations for the calibration problem for an increasing number of parameters, relative to the Adj TNC-0 algorithm. }
\begin{tabular}{|l|l|l|l|l|l|}
\hline
\# parameters & Adj TNC & Fin TNC &  \begin{tabular}{@{}c@{}}Fin TNC  \\relative evals. \end{tabular} & GA & \begin{tabular}{@{}c@{}}GA \\relative evals. \end{tabular}     \\ \hline
5             & 1098    & 1790 & 1.63   & 5256  & 4.79  \\ \hline
10            & 1200    & 3300  & 2.75  & 37715 & 31.4\\ \hline
15            & 1780    & 7139 & 4.01  & 135485 & 76.1\\ \hline
\end{tabular}
\end{table}

Assuming that a gradient based algorithm converges in approximately the same number of gradient/objective evaluations for either using finite differences or the adjoint method, the equivalent number of objective evaluations can be easily converted between the two methods. Namely, if using the adjoint method requires $g(m)$ total computational effort for computing the gradient, for some arbitrary function $g$, then finite differences will require $\frac{m+1}{4} g(m)$. So using finite differences will always scale an order of $m$ worse than the adjoint method. 
For the GA, it is clear that its cost is growing extremely fast with the number of parameters, and we only considered a platoon up to size 3 because of this reason. It was about 75 times as expensive as using TNC with the adjoint method for a problem size of only 15 parameters. 
We conclude that when the number of parameters becomes large, using the adjoint method with gradient based optimization can offer an arbitrarily large speed increase compared to either gradient free optimization or finite differences. 

\begin{figure}[h] 
\centering 
\includegraphics[ width=\textwidth]{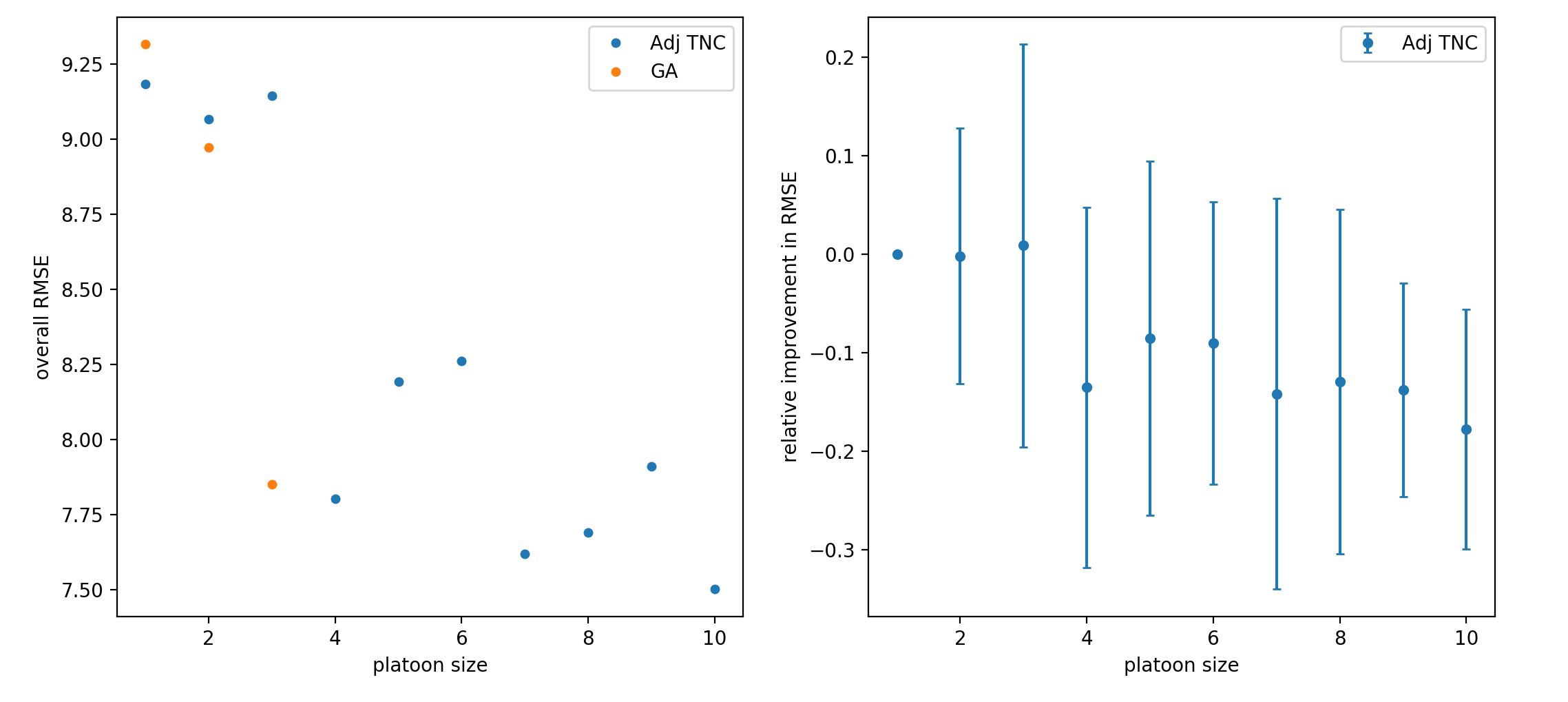} 
\caption{Left panel shows overall RMSE when using a varying platoon size for the calibration problem. Right panel shows the percentage improvement from using the larger platoon sizes compared to calibrating a single vehicle at a time. The bars show the standard deviation for each point. } \label{platacc}
\end{figure}
As for the effect of platoon size on the overall RMSE of all 100 vehicles, the results are shown in figure \ref{platacc}. Note overall RMSE is different than average RMSE because vehicles are weighted according to how many observations they have (see section \ref{implementation}). Also, the RMSE is expected to be higher here compared to the previous section since the calibration is being done sequentially using the leader's simulated trajectory instead of the leader's measured trajectory. The left panel shows the overall RMSE versus platoon size. The right panel shows the relative improvement in RMSE. To calculate the relative improvement, the percent change in RMSE compared to a platoon size of 1 is calculated for a specific platoon. Then the average and standard deviation of the percent change is calculated over all platoons of the specific size (e.g. 33 size 3 platoons or 10 size 10 platoons). The standard deviations are quite large: for any specific platoon, the improvement could be as large as 50\%, or the RMSE could sometimes be significantly worse. For a platoon size $n = 2$ or $3$, we saw a neglible overall improvement. In both those cases, we saw that the majority of platoons had an improvement of around $8\%-9\%$, but there were some platoons which achieved a significantly worse fit (20\%-40\%+ worse). These significantly worse fits suggest that in those cases, the algorithm was converging to bad local minima which apparently were avoided when calibrating vehicles one at a time. As the platoon size increased, the improvement became more consistent, and for platoon size $n = 9$ and $10$ we saw an overall $13.8\%$ and $17.8\%$ improvement respectively. The authors stress that this improvement was achieved without changing the algorithm or the model at all. The only thing that was changed was the platoon size ($n$ in Eq. \eqref{5}). 

Overall, these results suggest that considering a larger platoon size can improve calibration results, and further research on larger platoon sizes should be considered. In figure \ref{platacc} we saw that the GA starts to perform better than TNC when the platoon size increases. We also saw that for larger platoon sizes, there were some platoons which ended up with a significantly worse fit compared to the case of $n=1$. This suggests that the local search algorithm (TNC) was having problems converging to bad local minima for $n>1$. We therefore conclude that while using multiple guesses was adequate for the calibration of a single vehicle, a more sophisticated strategy is needed when multiple vehicles are being calibrated.

\section{Conclusion}\label{conclusion}
We considered an optimization based formulation of the calibration problem for car following models being calibrated to trajectory data. It was shown how to apply the adjoint method to derive the gradient or hessian for an arbitrary car following model formulated as either an ordinary or delay differential equation. Several algorithms for solving the calibration problem were compared, and it was found that the best overall algorithms were the genetic algorithm, l-bfgs-b, and truncated newton conjugate. 

For the calibration of a single vehicle at a time, it was found that using the adjoint method and a quasi-newton method gives slightly better performance than a genetic algorithm, and is 5 times faster. As the number of parameters increases, the speed increase offered by the adjoint method and gradient based optimization can become arbitrarily large (for 15 parameters, a truncated newton algorithm using the adjoint method was 75 times faster than a genetic algorithm). The adjoint method will always perform better than finite differences as the number of parameters becomes large, because computing the gradient with the adjoint method has a flat cost with respect to the number of parameters, whereas using finite differences has a cost which increases linearally with the number of parameters. Numerical experiments used the reconstructed NGsim data and the optimal velocity model.  

There are two main directions for future research. In this paper we consider only the calibration of a car following model, while a full microsimulator has several other important modules, such as route choice or lane changing decision models. The question of how to apply the adjoint method to those other components can be considered in the future so that the methodology could be applied to a full microsimulation model. There are also more questions regarding the calibration of car following models. Future research can consider the effect of a larger platoon size on the calibration problem, so that multiple vehicles (which share leaders) are calibrated at a time. Another interesting question is using the new highly accurate vehicle trajectory data \citep{66} to compare and validate different car following models. Since it was found during this study that lane changing vehicle trajectories are not described well by car following models, another question is how to incorporate lane changing dynamics into an arbitrary car following model, and the authors have recently proposed a method for doing so in \cite{100}. 

 \section*{Appendix A - Proof of Theorem 1 and Corollary 1}
  \begin{definition}%{Piecewise continuous.} 
We say that the function $h(y(t), z(t))$ is piecewise continuous on an interval $\mathcal{I}$ if there are a finite number of points $t \in \mathcal{I}$ where $h(y(t), z(t))$ is not continuous.
\end{definition}
\begin{definition}%{Existence and Uniqueness Theorem for ODEs.} 
(Existence and Uniqueness Theorem.) Define the initial value problem 
\begin{align*} 
\dot x(t) = h(t, x(t)), \quad x(t_0) = x_0
\end{align*}
where $h(t,y(t))$ is a continuous function on the region $\mathcal{R} = \{ (t,x) :  | t- t_0 | < s, |x - x_0 | < y \}$, 
and $h$ is uniformly Lipschitz in $x$, meaning 
\begin{align*} 
\left| \dfrac{\partial h}{\partial x} \right| \leq C
\end{align*}
for some constant $C$ in the region $\mathcal{R}$. Then the solution $x(t)$ exists and is unique on the interval $|t - t_0| < s*$ for some $0 < s^* < s$.
\end{definition}
 \begin{prf}[Proof of Theorem 1.]
We say that there is a discontinuity for vehicle $i$, either in its car following model or its adjoint variable, if at time $t$: a) its lead vehicle changes, b) its model regime changes, c) the vehicle changes from the car following model to the downstream boundary condition, or d) the loss function $f$ is not continuous at time $t$. We will refer to these different causes of discontinuities as types a) - d) respectively. Define the times $\theta_j$ for $j = 0, \ldots, m$ as the set of all times when when any vehicle $x_i, i = 1, \ldots, n$ experiences a discontinuity. The set $\{ \theta_j \}$ is of measure zero because each vehicle can contribute only a finite number of discontinuities. The assumption that $f$ and $h_i$ are piecewise continuous and switch regimes only a finite number of times means a vehicle can contribute only a finite number of type b) or d) discontinuities. A vehicle switches to the downstream boundary only once, contributing a single type c) discontinuity. And based on the physical nature of the problem it can be assumed that a vehicle can change its lead vehicle only a finite number of times in a finite interval. \\
Some of the times $\theta_j$ may depend on the switching condition $g$ in Eq. \eqref{regime}. These times may therefore depend on the parameters $p$, and we use the notation $\theta_j(p)$ to refer to the switching times for parameters $p$. Besides the set $\{ \theta_j \}$ being finite, we also require that 
\begin{align*} 
\underset{\epsilon \rightarrow 0}{\lim} \{\theta_j(p+\epsilon)\} = \{ \theta_j (p)\} \tag{17}\label{condition}
\end{align*}
meaning that the times of discontinuities must be continuous with respect to the parameters. \\
Let the set $\{ \theta_j \}$ be ordered in increasing time,and let it include the first and last times of the simulation $\theta_0 = \min \{t_1, \ldots, t_n\}$ and $\theta_m = \max \{T_{1}, \ldots, T_{n} \}$. Since all vehicles have the same lead vehicle on each time interval $[\theta_j, \theta_{j+1}]$, on that interval there exists some ordered set of vehicles $\chi_j$ which contains all vehicles which are simulated in that time interval, ordered such that the first vehicle has no simulated vehicles as leaders (i.e. the vehicle farthest downstream) and the last vehicle has no simulated vehicles as followers (i.e. the vehicle farthest upstream). Following the definition it is true that for $i \in \chi_j$ and $t \leq T_{i-1}$ all the partial derivatives $\partial f / \partial x_i$, $\partial h_i / \partial x_i$, $\partial h_i / \partial p$, $\partial h_i/ \partial x_{L(i)}$ as well as $f$ and $h_i$ are continuous with respect to their arguments, except possibly at the endpoints. \\
\begin{figure}[H] 
\centering 
\includegraphics[ width=.8\textwidth]{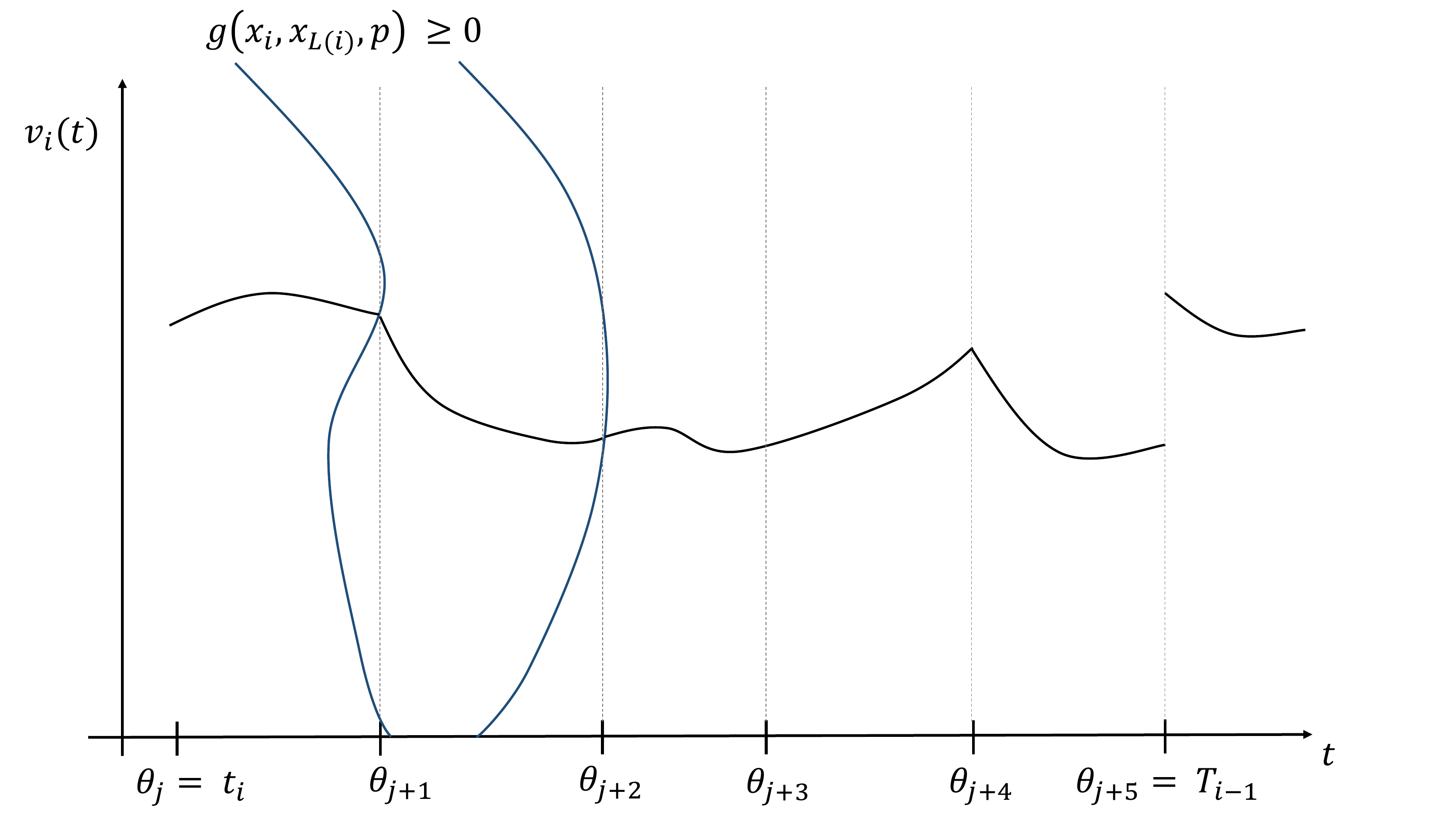} 
\caption{Illustration of the velocity $v_i(t)$ when $h_i(\cdot)$ may have discontinuities. The times $\theta_j$, where discontinuities can occur are marked as dashed lines. The region inside the two blue lines is the region where the switching condition $g(x_i, x_{L(i)}, p) \geq 0$, so the model changes regimes at times $\theta_{j+1}$ and $\theta_{j+2}$. The model is discontinuous at $\theta_{j+1}$ and $\theta_{j+2}$ but $v_i(t)$ is still continuous. At $\theta_{j+3}$ some other vehicle in the platoon experiences a discontinuity, but $i$ is unaffected. At $\theta_{j+4}$, the lead vehicle changes, causing a discontinuity in the model (but $v_i$ is still continuous at that time). At $\theta_{j+5}$, the vehicle switches to the downstream boundary condition which causes a jump discontinuity in $v_i$. } 
\end{figure}
First we show that $x_i(t), i \in \chi_0$ is continuous on the interval $[\theta_0, \theta_1]$. This is guaranteed by applying the existence and uniqueness theorem in order to each vehicle. Since by construction $h_i(x_i, x_{L(i)},p_i)$ is continuous in this interval excluding the right endpoint, and $\partial h_i / \partial x_i$ is bounded, then the initial value problem $\dot x_i = h_i$ with initial condition $x_i(t_i) = \hat x_i(t_i)$ has a solution, namely
\begin{align*} 
x_i(t) = \hat x_i(t_i) + \int_{t_i}^{\theta_1} h_i(x_i, x_{L(i)}, p_i) dt, \ \ t_i \leq t < \theta_1
\end{align*}
which is valid for $t_i \leq t < \theta_1$. Any discontinuity which may exist at $\theta_1$ is not felt by the integral since the point has measure 0. Since we solve for each vehicle following the order $\chi_0$, the continuity of $x_{L(i)}$ for $L(i) \in \chi_0$ is already obtained by the time it is needed to compute its follower. \\
For some arbitrary interval $[\theta_j, \theta_{j+1}]$, where the previous interval is already solved, take the left sided limit of the previous interval as the initial condition, i.e. $x_i(\theta_j) = \underset{\epsilon \rightarrow 0}{\lim} \ x_i(\theta_j-\epsilon)$. This point will always be defined by the previous interval. Then the solution on the new interval is defined as
\begin{align*} 
& x_i(t) = x_i(\theta_j) + \int_{\theta_j}^{\theta_{j+1}}h_i(x_i, x_{L(i)}, p_i) dt, \ \ \theta_j \leq t < \theta_{j+1} \\
& x_i(\theta_j) = \underset{\epsilon \rightarrow 0}{\lim} \ x_i(\theta_j-\epsilon)
\end{align*}
Note that any possible discontinuities in $h_i$ at times $\theta_j$ and $\theta_{j+1}$ do not affect the integral because they are of measure 0. Applying the existence and uniqueness theorem again, we have that $x_i(t)$ is continuous on $[\theta_j, \theta_{j+1}]$. Our choice of initial conditions guarentees continuity on the larger interval $[\theta_{j-1}, \theta_{j+1}]$. Thus it is clear from an induction argument that $x_i(t)$ will be continuous on the interval $[t_i, \theta_{j+1}]$, despite any possible discontinuities in $h_i$. \\
The exception to the continuity of $x_i$ is at time $T_{i-1}$, where the downstream boundary conditions will cause a jump discontinuity. Upon reaching the $\theta_j = T_{i-1}$, the remaining trajectory of $x_i$ is to be taken as 
\begin{align*} 
& x_i(t) = x_i(T_{i-1}) + \int_{T_{i-1}}^{T_i} \dot{\hat{x}}_i(t) dt \\
& x_i(T_{i-1}) = (\underset{\epsilon \rightarrow 0}{\lim} \ x_i^*(T_{i-1}-\epsilon) , \  \hat v_i(T_{i-1}) )^T
\end{align*} 
which guarantees continuity only in position (i.e. $x_i^*$, the first component of vector $x_i$) and the speed $v_i$ will have a jump discontinuity. \\
Thus we proven that the simulated trajectories $x_i$ are continuous on the interval $[t_i, T_{i-1}]$ (excluding $T_{i-1}$) by repeatedly applying the existence and uniqueness theorem on the intervals $[\theta_j, \theta_{j+1}]$ for an arbitrary $j$. \\
Then for the objective 
\begin{align*} 
F = \sum_{i=1}^n \int_{t_i}^{T_{i-1}}f(x_i, \hat x_i, t) dt
\end{align*}
since $f$ is assumed to have a finite number of discontinuities in $t$, $\hat x_i$ is assumed to be continuous, and we have shown $x_i(t)$ is continuous on the interval $t_i \leq t < T_{i-1}$, it follows that $F$ is continuous. \\
To show that $\lambda_i(t)$ is continuous, apply the existence and uniqueness theorem starting with the last interval $[\theta_{m}, \theta_{m-1}]$, going backwards over vehicles on $\chi_{m-1}$. Define 
\begin{align*} 
& \gamma_i = \mathbbm{1}(t \leq T_{i-1}) \left( \dfrac{\partial f}{\partial x_i} - \lambda_i^T(t)\dfrac{\partial h_i}{\partial x_i} \right) - \mathbbm{1}(G(i,t) \neq 0) \lambda_{G(i)}^T(t) \dfrac{\partial h_{G(i)}}{\partial x_i}
\end{align*}
and for an arbitrary interval $[\theta_{j}, \theta_{j-1}]$ define 
\begin{align*} 
& \lambda_i(t) = \lambda_i(\theta_j) + \int_{\theta_j}^{\theta_{j-1}} \gamma_i dt, \ \ \theta_j \geq t > \theta_{j-1} \\
& \lambda_i(\theta_j) = \underset{\epsilon \rightarrow 0}{\lim} \ \lambda_i(\theta_j + \epsilon)
\end{align*}
where additionally for the first interval for $\lambda_i$ (the interval where $\theta_j = T_i$), the initial condition should be $\lambda_i(\theta_j) = 0$. The justification for why $\lambda_i(t)$ is continuous follows the same argument as for $x_i$. The Lipschitz condition for $\lambda_i$ necessary to apply the existence and uniqueness theorem for $\lambda_i(t)$ is the same as the Lipschitz condition for $x_i$ ($\partial h_i/ \partial x_i$ is bounded). Additionally, by assumption we have all the partial derivatives of $h_i$ are continuous on each interval excluding the endpoints. Since the adjoint system contains the term $\mathbbm{1}(G(i,t) \neq 0) \lambda_{G(i)}^T(t) \dfrac{\partial h_{G(i)}}{\partial x_i}$ the $\lambda_i$ need to be solved backwards in the order $\chi_j$ so that the follower's adjoint variable is already proven as continuous by the time it is needed. Also note that $\dfrac{\partial h_{G(i)}}{\partial x_i}$ is simply $\partial h_k / \partial x_{L(k)}$ where $k = G(i)$.  \\
Because $\lambda_i$ is continuous on $[t_i, T_i]$, $x_i$ is continuous on $[t_i, T_{i-1})$, and $\partial h_i / \partial p$ contains a finite number of discontinuities, it follows that $\lambda_i^T(t) \partial h_i / \partial p$ is a piecewise continuous function with a finite number of discontinuities. Then the gradient
\begin{align*} 
\dfrac{d}{dp} F = \sum_{i=1}^n\left( -\int_{t_i}^{T_{i-1}} \lambda_i^T(t) \dfrac{\partial h_i}{\partial p} dt \right)
\end{align*}
is continuous since the number of discontinuities over the integral is of measure 0. \qed \\
\end{prf}
\begin{prf}[Proof of Corrolary 1.]
Showing $F$ is lipschitz continuous is equivalent to showing that $d F / dp$ is bounded. To do this define 
\begin{align*} 
& \alpha(t) = \int_{T_i}^{t} \left| \mathbbm{1}(t \leq T_{i-1}) \dfrac{\partial f}{\partial x_i} \right| +\left| \mathbbm{1}(G(i,t) \neq 0) \lambda_{G(i)}^T(t) \dfrac{\partial h_{G(i)}}{\partial x_i} \right| dt \\
& \beta(t) = \dfrac{\partial h_i}{\partial x_i}
\end{align*}
Then by definition we have
\begin{align*} 
\lambda_i(t) \leq \alpha(t) + \int_{T_i}^t \beta(s) \lambda_i(s) ds
\end{align*}
and by Gronwall's inequality in integral form
\begin{align*} 
\lambda_i(t) \leq \alpha(t) \exp \left( \int_{T_i}^t\beta(s) ds \right) \tag{18}\label{gronwall}
\end{align*}
To show that all $\lambda_i(t)$ are bounded, prove that each $\lambda_i , i \in \chi_j$ are bounded in the interval $[\theta_j, \theta_{j-1}]$ going backwards over the order specified by $\chi_{j-1}$, starting with $j=m$. Since we assumed that $\partial f / \partial x_i$ and $\partial h_i / \partial x_i$ are both bounded, and for the last vehicle $k$ in $\chi_j$, $G(i,t) = 0$, then it follows from Eq. \eqref{gronwall} that $\lambda_k(t)$ is bounded in the interval $[\theta_j, \theta_{j-1}]$. Then after showing this for $k$, it can be shown for the other vehicles in $\chi_j$. Repeating this process for each interval shows that each $\lambda_i(t)$ is bounded on $[T_i, t_i]$. \\
Then for the gradient Eq. \eqref{16} since we showed $\lambda_i(t)$ is bounded, and $\partial h_i / \partial p_i$ is bounded by assumption, then it follows that $d F / dp$ is also bounded, and hence $F$ is Lipschitz continuous. \qed
\end{prf} 
 \subsection*{A.1 - Example with OVM}
 Written as a first order model OVM is 
 \begin{align*} 
 \dot x_i(t) = \dfrac{d}{dt} \bpm x_i^*(t) \\ v_i(t) \epm = \bpm v \\ c_4c_1 \tanh ( c_2 s_i(t) - c_3 - c_5) - c_4c_1 \tanh ( - c_3) - c_4 v_i(t) \epm = h_i(x_i, x_{L(i)}, p_i) 
 \end{align*}
Where $p_i = (c_1, c_2, c_3, c_4, c_5) > \textbf{0}$ and $s_i(t) = x_i^*(t) - x_{L(i)}(t) - l$ (where $l$ is the length of the lead vehicle). Since it is a single regime model, it will be everwhere continuous. The partial derivative 
\begin{align*} 
\dfrac{\partial h_i}{\partial x_i} = \bpm 0 & 1 \\ -c_1c_2c_4 \sech ( c_3 + c_5 - c_2(s_i))^2 & -c_4 \epm \Rightarrow \left| \dfrac{\partial h_i}{\partial x_i} \right| \leq \bpm 0 & 1 \\ c_1c_2c_4 & c_4 \epm 
\end{align*}
so it can be bounded by a constant for an arbitrary vector space. The other partial derivatives can be bounded as
\begin{align*} 
\left| \dfrac{\partial h_i}{\partial x_{L(i)}} \right|  \leq  \bpm 0 & 0 \\ c_1c_2c_4 & 0 \epm \\
\left| \dfrac{\partial h_i}{\partial p_i} \right| \leq \bpm 0, & 0, & 0, & 0, & 0 \\ 2c_4, & c_1 c_4 s_i, & c_1c_4, & v_i + 2 c_1, & c_1c_4  \epm
\end{align*}
And so the objective will be Lipschitz continuous assuming that the headway $s_i(t)$ and velocity $v_i(t)$ are finite.
 \section*{Appendix B - Supplemental figures/tables for comparison of algorithms}
 
\begin{table}[H] %i don't know why it stays at the bottom of the page, none of the options will make it go to the top of the page. 
\centering
\caption{The full table showing all variants of the 7 algorithms considered.}
\begin{tabular}{|l|l|l|l|l|l|l|l|l|}
\hline
Algorithm & \begin{tabular}{@{}c@{}}\% found  \\global opt\end{tabular} & \begin{tabular}{@{}c@{}} Avg.  \\ RMSE (ft) \end{tabular} &  \begin{tabular}{@{}c@{}} Avg. \\  Time (s)\end{tabular} & \begin{tabular}{@{}c@{}} Avg. RMSE / \% \\  over global opt\end{tabular} &\begin{tabular}{@{}c@{}}Initial  \\ Guesses\end{tabular} & \# Obj. & \# Grad.& \# Hess. \\ \hline
Adj BFGS-0 & 94.0  & 6.46& 7.42 & .10 / 2.0\% &3 &307.1 & 307.1 & 0 \\ \hline
Adj BFGS-7.5 &85.2  &6.56 &4.16 & .20/6.8\% & 1.67 & 165.0&  165.0& 0 \\ \hline
Adj BFGS-$\infty$& 75.6& 6.96& 2.55 & .59 / 15.4\% & 1 & 107.3 &107.3 & 0\\ \hline
GA & 95.0 & 6.47 &33.5 & .10/2.2\%& -  & 5391 & 0& 0\\ \hline
NM & 39.4 & 7.25 & 6.85   & .88/21.0\%& 1 & 1037 & 0 &0 \\ \hline
Fin BFGS-0& 94.3 & 6.46 & 11.4  & .10/2.2\%&3 & 311.5 &311.5 &0 \\ \hline
Fin BFGS-7.5&85.7 & 6.55 & 6.32  & .19/6.3\%&1.66 & 164.8& 164.8& 0\\ \hline
Fin BFGS-$\infty$& 77.0& 6.88 & 3.91 & .52/13.8\%& 1& 107.3& 107.3& 0\\ \hline
Adj GD-0&20.2 & 7.58 & 25.5  & 1.21/28.2\%& 3 & 1374 & 542.5 & 0\\ \hline
Adj GD-7.5& 13.4 & 7.83 & 18.1   & 1.46/40.8\%& 1.93 & 1094 & 436.6 &0 \\ \hline
Adj GD-Inf& 8.8& 8.90 & 8.75 & 2.53/67.2\% & 1 & 719.3 &282.6 &0 \\ \hline
Adj SQP-0& 29.1 & 8.37 &21.1   & 2.01/61.7\%&3 &287.6 & 83.8 & 80.8\\ \hline
Adj SQP-7.5&21.5  & 8.59 & 15.4 & 2.22/71.4\%& 2.17 & 244.8 & 71.3 & 69.1 \\ \hline
Adj SQP-$\infty$& 10.0  &11.1 & 5.87 & 4.70/140.7\% & 1 & 141.6 & 39.4 &  38.4 \\ \hline
SPSA& 4.2  & 53.2  & 38.5  & 46.8/786\%& 1 & 6001 & 0 &0  \\ \hline
Fin GD-$\infty$& 9.1 & 8.86 & 14.7  & 2.50/66.2\%& 1 & 657.3 & 258.5 &0 \\ \hline
Fin SQP-$\infty$& 2.6 & 9.61 & 15.8   & 3.25/81.8\%&1 & 179.4 & 67.2 & 66.2 \\ \hline
Adj TNC-0& 90.7 & 6.43 & 6.87  & .05/2.4\%&3 & 285.7 & 285.7& 0 \\ \hline
Adj TNC-7.5& 82.7  & 6.48 & 3.82 & .11/5.0\%& 1.63 & 152.3 &152.3 &0 \\ \hline
Adj TNC-$\infty$& 77.3 &  6.62& 2.26 & .25/7.7\%& 1 & 94.1 & 94.1 & 0 \\ \hline
Fin TNC-0& 36.3 & 6.84 & 13.2 & .47/11.0\% & 3 & 302.4& 302.4& 0 \\ \hline
Fin TNC-7.5& 24.9  & 7.05 & 8.35 & .69/20.8\% & 1.82 & 183.2 & 183.2 & 0 \\ \hline
Fin TNC-$\infty$& 18.0 & 7.63 & 4.49  & 1.27/33.2\%& 1 & 100.8& 100.8& 0\\ \hline
\end{tabular}
\end{table}

\begin{figure}[H] 
\centering 
\includegraphics[ width=\textwidth]{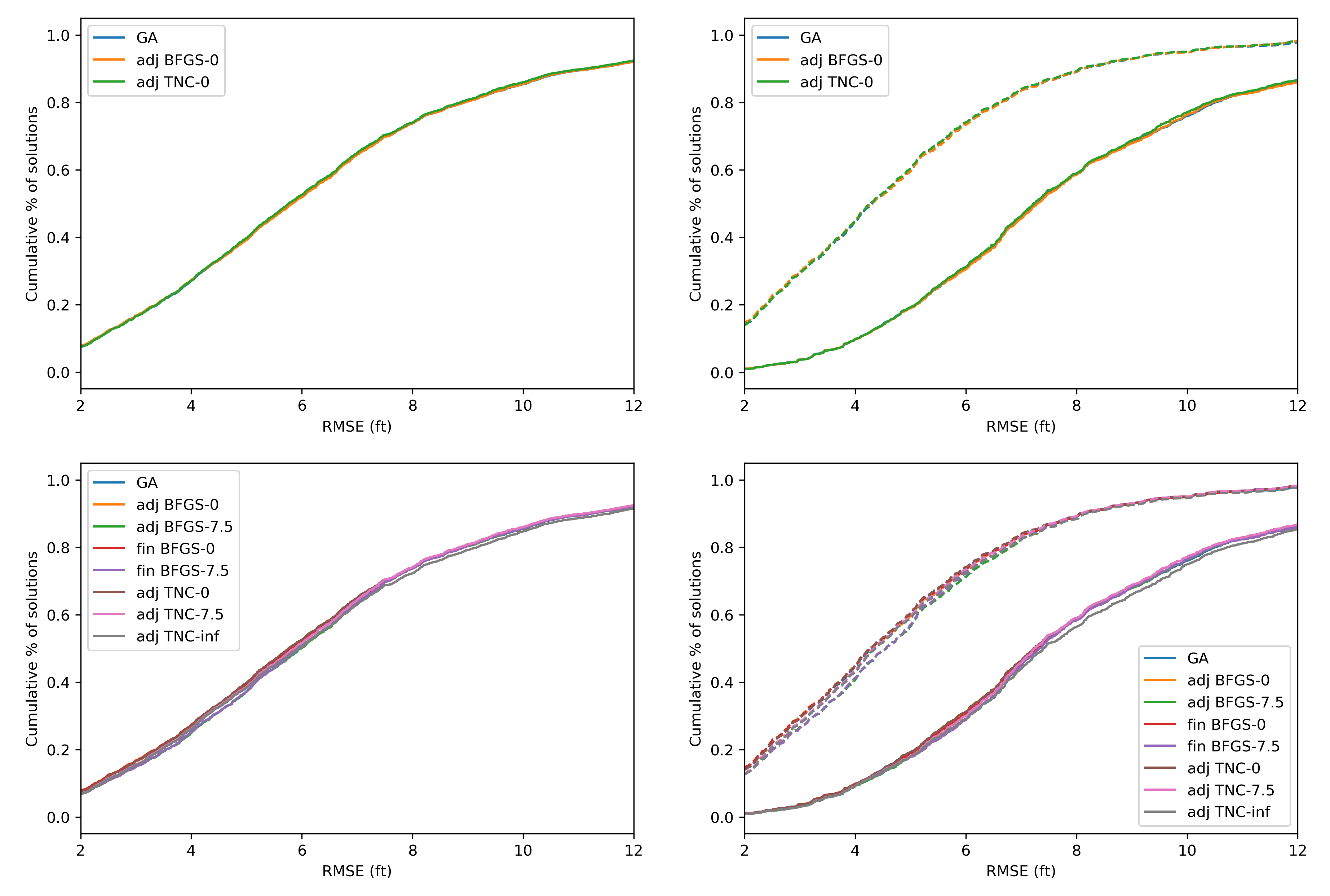} 
\caption{Shows the distribution of RMSE for different algorithms tested. Left panels show overall RMSE, right panels show the RMSE for vehicles that experience lane changes (solid) and those that don't (dashed). The RMSE for lane changing vehicles is much higher because car following models don't describe lane changes well.  For example, for the genetic algorithm, the average RMSE for vehicles which do not experience lane changes is 4.75, whereas it is 8.17 for vehicles which do. The other algorithms have very similar numbers. } 
\end{figure}

\section*{Appendix C - Adjoint method for gradient with DDE model}
\noindent We will now deal with calculating the gradient for Eq. \eqref{6}. As before, if not explicitly stated, quantities can be assumed to be evaluated at time $t$. Define the augmented objective function: 
\bg{ L = \sum_{i=1}^n \left( \int_{t_i + \tau_i}^{T_{i-1}^*} f(x_i, \hat x_i) + \lambda_i^T(t) \bigg( \dot x_i(t) - h_i(x_i(t), x_i(t-\tau_i), x_{L(i)}(t-\tau_i), p_i) \bigg) dt  + \int_{T_{i-1}^*}^{T_i} \lambda_i^T ( \dot x_i - \dot{\hat{x}}_i ) dt \right)   \tag{19}\label{27}}
and since $F = L$
\begin{align*}
&\dfrac{d}{dp} F = \sum_{i=1}^n \bigg[ \int_{t_i + \tau_i }^{T_{i-1}^* } \dfrac{\partial f}{\partial x_i}\dfrac{\partial x_i}{\partial p} - \lambda_i^T(t)\dfrac{\partial h_i}{\partial x_i}\dfrac{\partial x_i}{\partial p} - \lambda_i^T(t)\dfrac{\partial h_i}{\partial x_i(t-\tau_i)}\dfrac{\partial x_i(t- \tau_i)}{\partial p} 
- \lambda_i^T(t)\dfrac{\partial h_i}{\partial x_i(t-\tau_i)}\dot x_i(t-\tau_i) \dfrac{\partial (t-\tau_i)}{\partial p} \\ 
& - \lambda_i^T(t)\dfrac{\partial h_i}{\partial p} dt  - \int_{t_i + \tau_i}^{T_i} \dot \lambda_i^T(t)\dfrac{\partial x_i}{\partial p} dt \bigg] 
+ \sum_{i=1}^{n} \sum_{j \in S(i)}\int_{t_{i}+\tau_j}^{T_{i}+\tau_j}\mathbbm{1}(G(i,t-\tau_j) = j)\bigg[ + \ldots \\ & 
 -\lambda_{j}^T(t) \dfrac{ \partial h_{j}}{\partial x_i(t-\tau_{j})}\dfrac{\partial x_i(t-\tau_{j}) }{\partial p} - 
\lambda_{j}^T(t)\dfrac{\partial h_{j}}{\partial x_i(t - \tau_{j})}\dot{x}_i(t-\tau_{j})\dfrac{\partial(t - \tau_{j})}{\partial p} dt\bigg] \tag{20}\label{28} 
\end{align*}
where $S(i)$ is defined as the set of all followers for vehicle $i$ and we have chosen $\lambda_i^T(T_{i-1}^*) = 0$. For a DDE, use a change of coordinates on terms containing $\partial x_i(t - \tau) / \partial p $. 
\begin{align*}
&-\int_{t_i+\tau_i}^{T_{i-1}^*}  \lambda_i^T(t)\dfrac{\partial h_i}{\partial x_i(t-\tau_i)}\dfrac{\partial x_i(t- \tau_i)}{\partial p_i}dt 
= -\int_{t_i+\tau_i}^{T_{i-1}^* - \tau_i} \lambda_i^T(t+\tau_i) \dfrac{\partial h_i(t + \tau_i)}{\partial x_i(t- \tau_i)}\dfrac{ \partial x_i(t)}{\partial p_i}dt  \\
& \sum_{i=1}^{n} \sum_{j \in S(i)}\int_{t_{i}+\tau_j}^{T_{i}+\tau_j}-\mathbbm{1}(G(i,t-\tau_j) = j) \lambda_{j}^T(t) \dfrac{ \partial h_{j}}{\partial x_i(t-\tau_{j})}\dfrac{\partial x_i(t-\tau_{j}) }{\partial p}dt \\ 
& = \sum_{i=1}^{n} \sum_{j \in S(i)}\int_{t_{i}+\tau_{i}}^{T_{i}}-\mathbbm{1}(G(i,t) = j) \lambda_{j}^T(t+\tau_j) \dfrac{ \partial h_{j}(t+\tau_j)}{\partial x_i(t-\tau_{j})}\dfrac{\partial x_i(t) }{\partial p}dt  \tag{21}\label{29}
\end{align*}
Where we make use of the fact that $\partial x_i (t) / \partial p  = 0$ for $t \in [t_i, t_i+\tau_i]$ since the initial history function doesn't depend on the parameters. Now substituting Eq. $\eqref{29}$ into Eq. $\eqref{28}$ and grouping terms, one obtains the adjoint system: 
\begin{align*}
& - \dot \lambda_i^T(t) + \mathbbm{1}(t \leq T_{i-1}^*)\left( \dfrac{\partial f}{\partial x_i} - \lambda_i^T(t)\dfrac{\partial h_i}{\partial x_i} \right) - \mathbbm{1}(t \leq T_{i-1}^* - \tau_i) \lambda_i^T(t+\tau_i)\dfrac{\partial h_i(t + \tau_i)}{\partial x_i(t-\tau_i)} \ldots \\ 
&  - \sum_{j \in S(i)}\mathbbm{1}( G(i,t) = j)\lambda_{j}^T(t+\tau_{j})\dfrac{ \partial h_{j}(t+\tau_{j})}{\partial x_i(t - \tau_{j})}        = 0, \quad  t \in [T_i, t_i + \tau_i], \ \ \ \forall i    \tag{22}\label{32}
\end{align*}   
\noindent with initial conditions $\lambda_i(T_{i}) = 0$. For a DDE model, the adjoint system also becomes a DDE. After solving for the adjoint variables, one can compute the gradient: 
\begin{align*} 
&\dfrac{d}{dp} F = \sum_{i=1}^n \bigg[ \int_{t_i+\tau_i}^{T_{i-1}^*} - \lambda_i^T(t)\dfrac{\partial h_i}{\partial x_i(t - \tau_i)} \dot x_i(t - \tau_i)\dfrac{\partial (t - \tau_i)}{\partial p} -\lambda_i^T\dfrac{\partial h_i}{\partial p}dt \bigg] \\
& + \sum_{i=1}^{n} \sum_{j \in S(i)}\int_{t_{i}+\tau_j}^{T_{i}+\tau_j}-\mathbbm{1}(G(i,t-\tau_j) = j)\lambda_{j}^T(t)\dfrac{\partial h_{j}}{\partial x_i(t - \tau_{j})}\dot{x}_i(t-\tau_{j})\dfrac{\partial(t - \tau_{j})}{\partial p} dt 
 \tag{23}\label{31}
\end{align*} 

 \section*{Appendix D - Extension of adjoint method for Hessian}
  In this section, for a scalar being differentiated with respect to two vectors, the result is a matrix where the rows correspond to the first vector, and the columns correspond to the second vector. For example, 
 \begin{align*} 
\dfrac{ \partial f }{\partial x_i \partial p} = \bpm \partial f / \partial x_i^* \partial p^1 & \partial f / \partial x_i^* \partial p^2 & \ldots & \partial f / \partial x_i^* \partial p^m \\ 
 \partial f / \partial v_i \partial p^1 & \partial f / \partial v_i \partial p^2 & \ldots & \partial f / \partial v_i \partial p^m \epm 
 \end{align*}
 For a matrix being differentiated to two vectors, the result is a 3-dimensional array which should be interpreted as a vector of matrices. For example 
 \begin{align*} 
 \dfrac{\partial h_i}{\partial p \partial x_i } = \bpm \partial h_i^1 / \partial p \partial x_i \\ \partial h_i^2 / \partial  p \partial x_i \epm
 \end{align*}
 where $h_i^1$ and $h_i^2$ refer to the two components of $h_i$. So the normal rules of linear algebra can be applied, e.g. 
 \begin{align*} 
& \dfrac{\partial x_i}{\partial p}^T \dfrac{\partial h_i}{\partial p \partial x_i}^T \lambda_i = \dfrac{\partial x_i}{\partial p}^T \bpm \dfrac{h_i^1}{\partial p \partial x_i}^T,  & \dfrac{h_i^2}{\partial p \partial x_i}^T \epm \lambda_i = \bpm \dfrac{\partial x_i}{\partial p}^T \dfrac{\partial h_i^1}{\partial x_i \partial p},  & \dfrac{\partial x_i}{\partial p}^T \dfrac{\partial h_i^2}{\partial x_i \partial p} \epm \bpm \lambda_i^1 \\ \lambda_i^2 \epm \\
& =  \dfrac{\partial x_i}{\partial p}^T \dfrac{\partial h_i^1}{\partial x_i \partial p}\lambda_i^1 + \dfrac{\partial x_i}{\partial p}^T \dfrac{\partial h_i^2}{\partial x_i \partial p} \lambda_i^2
 \end{align*}
 where $\lambda_i^1$ and $\lambda_i^2$ are the two components of the adjoint variable. 
 
 \noindent The adjoint method extends in a straightforward way for computing the Hessian. Using finite differences would have complexity $\mathcal{O}(m^2 T(n))$, whereas the adjoint method can be shown to have complexity $\mathcal{O}(m T(n))$ \citep{30}. \\
 When calculating the gradient, the benefit of the adjoint method is that you avoid calculating $\partial x_i / \partial p$ terms. When calculating the Hessian, the adjoint method will be used to avoid calculating $\partial ^2 x_i/ \partial p^2$ terms. We will still be left with the unknowns $\partial x_i / \partial p$, which need to be calculated using the ``variational" approach, also known as the ``forward sensitivity" approach (see \cite{18, 22}).  

To obtain the variational system, differentiate the model Eq. \eqref{4} with respect to the parameters 
\[ \dfrac{d}{dp} \dot x_i = \dfrac{d}{dp} h_i(x_i, x_{L(i)}, p_i)  \]
and then switch the order of differentiation on the left hand side to obtain the variational system 
\begin{align*} 
& \dfrac{d}{dt} \dfrac{\partial x_i}{\partial p} = \dfrac{\partial h_i}{\partial x_i}\dfrac{\partial x_i}{\partial p} + \dfrac{ \partial h_i}{\partial p_i} + \mathbbm{1}(L(i,t) \in [1, n]) \dfrac{\partial h_i}{\partial x_{L(i)}}\dfrac{\partial x_{L(i)}}{\partial p} , \quad t \in [t_i, T_{i-1}] \ \ \ \forall i . \tag{24}\label{24}
\end{align*}
Eq. $\eqref{24}$ gives ${\partial x_i(t)}/{\partial p}$ starting from the initial conditions ${\partial x_i(t_i)}/{\partial p}$. Note that the variational system for $x_i$ is only defined up to time $T_{i-1}$ because $\partial x_i (t)/ \partial p = \partial x_i (T_{i-1})/ \partial p$ for $t \in [T_{i-1}, T_i]$. Now starting from Eq. \eqref{10}, differentiate twice instead of once
\begin{align*} 
&\dfrac{d^2}{d p^2} F = \dfrac{d^2}{d p^2} L = \sum_{i=1}^n\bigg[ \int_{t_i}^{T_{i-1}}\dfrac{\partial x_i}{\partial p}^T \dfrac{\partial ^2 f}{\partial x_i^2}^T\dfrac{\partial x_i}{\partial p} + \dfrac{\partial f}{\partial x_i}\dfrac{\partial^2 x_i}{\partial p^2} - \dfrac{\partial x_i}{\partial p}^T\dfrac{\partial^2 h_i}{\partial x_i^2}^T \lambda_i \dfrac{\partial x_i}{\partial p} - \lambda_i^T\dfrac{\partial h_i}{\partial x_i}\dfrac{\partial ^2 x_i}{\partial p^2} + \ldots \\
 &- \dfrac{\partial x_i}{\partial p}^T \dfrac{\partial^2 h_i}{\partial p \partial x_i}^T\lambda_i - \lambda_i^T \dfrac{\partial^2 h_i}{\partial p^2}  - \dfrac{\partial^2 h_i}{\partial x_i \partial p}^T\lambda_i \dfrac{\partial x_i}{\partial p} dt + \int _{t_i}^{T_i} \lambda_i^T\dfrac{\partial^2 \dot{x}_i}{\partial p^2}dt \bigg] +\ldots \\
&+ \sum_{i=1}^{n}\int_{t_{i}}^{T_i}  \mathbbm{1}(G(i,t) \neq 0)\bigg[ - \dfrac{\partial x_i}{\partial p}^T\dfrac{\partial^2 h_{G(i)}}{\partial x_i^2}^T\lambda_{G(i)} \dfrac{\partial x_i}{\partial p} - \dfrac{\partial x_i}{\partial p}^T\dfrac{\partial^2 h_{G(i)}}{\partial p \partial x_i}^T\lambda_{G(i)} - \dfrac{\partial x_i}{\partial p}^T\dfrac{\partial^2 h_{G(i)}}{\partial x_{G(i)}\partial x_i}^T \lambda_{G(i)}\dfrac{\partial x_{G(i)}}{\partial p} + \ldots \\ 
&- \lambda_{G(i)}^T\dfrac{\partial h_{G(i)}}{\partial x_i}\dfrac{\partial^2 x_i}{\partial p^2} - \dfrac{\partial^2 h_{G(i)}}{\partial x_i \partial p}^T \lambda_{G(i)}\dfrac{\partial x_i}{\partial p} - \dfrac{\partial x_{G(i)}}{\partial p}^T\dfrac{\partial^2 h_{G(i)}}{\partial x_i \partial x_{G(i)}}^T \lambda_{G(i)}\dfrac{\partial x_i}{\partial p} dt \bigg]
\end{align*}

The adjoint variables are defined the same way whether the gradient or Hessian is being computed. Eq. \eqref{16} gives the adjoint system for the unmodified objective $F$.  
After solving for all the adjoint variables, in addition to solving the variational system Eq. \eqref{24}, one can calculate the Hessian:
\begin{align*} 
& \dfrac{d^2}{d p^2} F = \sum_{i=1}^n\bigg[ \int_{t_i}^{T_{i-1}}\dfrac{\partial x_i}{\partial p}^T \dfrac{\partial ^2 f}{\partial x_i^2}^T\dfrac{\partial x_i}{\partial p}  - \dfrac{\partial x_i}{\partial p}^T\dfrac{\partial^2 h_i}{\partial x_i^2}^T \lambda_i \dfrac{\partial x_i}{\partial p}  - \dfrac{\partial x_i}{\partial p}^T \dfrac{\partial^2 h_i}{\partial p \partial x_i}\lambda_i - \lambda_i^T \dfrac{\partial^2 h_i}{\partial p^2}  +\ldots \\ 
&- \dfrac{\partial^2 h_i}{\partial x_i \partial p}^T\lambda_i \dfrac{\partial x_i}{\partial p} dt \bigg] 
+ \sum_{i=1}^{n}\int_{t_{i}}^{T_i}  \mathbbm{1}(G(i,t) \neq 0) \bigg[  - \dfrac{\partial x_i}{\partial p}^T\dfrac{\partial^2 h_{G(i)}}{\partial x_i^2}^T\lambda_{G(i)} \dfrac{\partial x_i}{\partial p}  -\dfrac{\partial x_i}{\partial p}^T\dfrac{\partial^2 h_{G(i)}}{\partial p \partial x_i}^T\lambda_{G(i)} + \ldots \\
&- \dfrac{\partial x_i}{\partial p}^T\dfrac{\partial^2 h_{G(i)}}{\partial x_{G(i)}\partial x_i}^T \lambda_{G(i)}\dfrac{\partial x_{G(i)}}{\partial p} - \dfrac{\partial^2 h_{G(i)}}{\partial x_i \partial p}^T \lambda_{G(i)}\dfrac{\partial x_i}{\partial p} - \dfrac{\partial x_{G(i)}}{\partial p}^T\dfrac{\partial^2 h_{G(i)}}{\partial x_i \partial x_{G(i)}}^T \lambda_{G(i)}\dfrac{\partial x_i}{\partial p} dt \bigg] + \ldots  \tag{25}\label{25}
\end{align*}

%\nocite{*} %for bibtex
\bibliographystyle{unsrt}
\bibliography{mysources1}

\end{document}